\documentclass[traditabstract]{aa} % for the abstract without structuration 
\usepackage{graphicx}
\usepackage{txfonts}
\usepackage{natbib}
\usepackage{url}
\usepackage{epsfig}
\usepackage{lscape}

\begin{document}
   \title{Photometric identification of blue horizontal branch stars \footnotemark}

   \author{K.W. Smith
          \inst{1}
          \and
          C.A.L. Bailer-Jones
          \inst{1}
          \and
          R.J. Klement
          \inst{1}
          \and
          X.X. Xue
          \inst{2}
          }

   \institute{Max Planck Institute for Astronomy (MPIA),
              K\"onigstuhl 17, Heidelberg 69117 , Germany.
              \email{smith@mpia-hd.mpg.de}
         \and
The National Astronomical Observatories, 
CAS, 20A Datun Road, Chaoyang District, 
100012, Beijing, China }

   \date{Received 9.03.2010; accepted 26.7.2010}

% \abstract{}{}{}{}{} 
% 5 {} token are mandatory
 
   \abstract {

     % Blue Horizontal Branch stars are useful structure tracers because they
     % have well defined luminosities. They are also bright, and therefore can
     % be used to probe greater distances than some other tracers. They can be
     % identified photometrically with some success, but simple colour cuts
     % yield samples that are significantly contaminated. Spectroscopic
     % classification is more accurate but is available for limited numbers
     % and brihter sources.

% below, the original
%

     We investigate the performance of some common machine learning
       techniques in identifying Blue Horizontal Branch (BHB) stars from
       photometric data.  To train the machine learning algorithms, we use
       previously published spectroscopic identifications of BHB stars from
       Sloan Digital Sky Survey (SDSS) data. We investigate the performance
     of three different techniques, namely k nearest neighbour classification,
     kernel density estimation for discriminant analysis and a support vector
     machine (SVM).  We discuss the performance of the methods in terms of both
     completeness (what fraction of input BHB stars are successfully returned
     as BHB stars) and contamination (what fraction of contaminating sources
     end up in the output BHB sample). We discuss the prospect of trading off
     these values, achieving lower contamination at the expense of lower
     completeness, by adjusting probability thresholds for the
     classification. We also discuss the role of prior probabilities in the
     classification performance, and we assess via simulations the reliability
     of the dataset used for training.  
Overall it seems that no-prior gives the best completeness, 
but adopting a prior lowers the contamination. We find that the support vector
       machine generally delivers the lowest contamination for a given level of
       completeness, and so is our method of choice. Finally, we classify a
     large sample of SDSS Data Release 7 (DR7) photometry using the SVM
     trained on the spectroscopic sample.  We identify 27,074 probable BHB
     stars out of a sample of 294,652 stars. We derive photometric parallaxes
     and demonstrate that our results are reasonable by comparing to known
     distances for a selection of globular clusters.  We attach our
     classifications, including probabilities, as an electronic table, so that
     they can be used either directly as a BHB star catalogue, or as priors to
     a spectroscopic or other classification method. We also provide our final
     models so that they can be directly applied to new data.

}
  % aims heading (mandatory)
{

  % In this paper, we use a spectroscopic classification of a large number of
  % Blue Horizontal Branch stars in the SDSS dataset to define a training set
  % in SDSS photometric colours, which we use to train machine learning
  % algorithms. We test several common machine learning algorithms against
  % each other and select one with the best performance. We then obtain a
  % large sample of new SDSS photometry and identify Blue Horizontal Branch
  % stars amongst this sample. We conduct a rudimentary distance test using
  % known globular clusters to demonstrate the viability of this
  % classification, and also some of its limitations.

}
  % methods heading (mandatory)
{

%   The three machine learning techniques we use are k-nearest neighbours,
%   Kernel density estimation for discriminant analysis, and support vector
%   machine for classification.

}
  % results heading (mandatory)
{

%   We found that the SVM and KDE provide the best performance, with the SVM
%   achieving marginally better results than the KDE. We therefore selected the
%   SVM as the classifier of choice. With this algorithm, we were able to 
% identify over 20,000 sources as probable BHB stars, including about 2000
% from the original spectroscopic sample.

}
  % conclusions heading (optional), leave it empty if necessary 
{

}

   \keywords{BHB stars --
                machine learning --
                classification
               }

   \maketitle
%
%________________________________________________________________

\renewcommand{\thefootnote}{$\star$}
\footnotetext[1]{Tables~\ref{catalogue},~\ref{oneClassData} and~\ref{twoClassData} are only available in electronic form
at the CDS via anonymous ftp to cdsarc.u-strasbg.fr (130.79.128.5)
or via http://cdsweb.u-strasbg.fr/cgi-bin/qcat?J/A+A/}

\section{Introduction}

The blue horizontal branch (BHB) stars are old, metal-poor halo stars. They
are of interest as tracers of Galactic structure because they are more
luminous than most giant branch or population II main sequence stars, have a
narrow range of intrinsic luminosities (hence 'horizontal branch') and display
spectral features rendering them identifiable, in particular a strong Balmer
jump and narrow strong Balmer lines.  There is therefore an interest in
building large, reliable samples of them, particularly in the context of
wide-field halo surveys such as the Sloan digital sky survey (SDSS) and the
forthcoming Pan-Starrs survey.  BHB stars are always of interest whenever halo
structure is studied due to their strength as distance indicators.  Recent
studies which have concentrated on BHB stars to trace structure include
\cite{harrigan}, who searched for moving groups in the halo, \cite{xue2} who
used them to search for close pairs, implying the existence of halo
substructure, \cite{kinman} who searched for a population of BHB stars
associated with the thick disk, and \cite{ruhland}, who investigated structure
in the Sagittarius dwarf and streams. 
The main problem with BHBs as tracers is their
relative sparseness compared to other tracers such as turnoff stars. This
means that large, pure samples are highly desirable for structure tracing
studies.

In this paper, we take as our lead several recent studies of BHB spectra from
SDSS/SEGUE and attempt to use the reliable and large samples of BHBs detected
as a training set to build models aimed at identifying BHBs from the
photometry alone. With this tool, we hope to be able to extend the available
sample of known (or better, strongly suspected) BHB stars from SDSS and other
surveys, with a view either to use our sample directly to trace structure, or
at least to guide follow-up studies with spectra.

The main three studies we follow are those of \cite{yanny}, who identified a
colour cut in the {\em u-g, g-r} colour-colour diagram that yields most of the
available BHB population, \cite{sirko} who used spectra to identify a reliable
sample of 700-1000 BHBs (the size of the sample depends on the $g$ magnitude
and the reliability desired), and most importantly \cite{xue}, who analysed a
sample of SDSS DR6 data using similar techniques to \citeauthor{sirko} and
extended the reliable list of BHBs to over 2,500 objects. The 
method of \citeauthor{xue} is discussed in more detail in 
Sect.~\ref{xueReanalysisSect}.

We have selected three machine learning methods to investigate.  These are a
k-Nearest Neighbour (kNN) technique, a kernel density estimator (KDE) and a
support vector machine (SVM). We also apply the decision boundary in
($u-g$,~$g-r$) colour space suggested by \cite{yanny} for comparison. After
the colour cut, the kNN method is probably the simplest algorithm we
consider. One example of its use can be found in \cite{marengo}.  Examples of
KDE use in classification problems in astrophysics include \cite{gao},
\cite{richards1}, \cite{richards2} and \cite{ruhland}. 
The SVM works by identifying a
decision boundary in a multidimensional space \citep{vapnik} (in this case the
space of SDSS colours) based on a training set containing examples of two or
more classes of object -- for our purposes BHB stars and non-BHB
contaminants. The SVM performance should be equivalent to that obtainable with
a neural network, but it has the advantage of being highly adaptable and
relatively easy to use. Its main drawback is its inability to provide genuine
probability estimates for classes, because it does not model the distribution
of the data. This is discussed further in Sect.~\ref{svms}.  SVMs have been
used on classification problems by various authors, for example
\cite{tsalmantza2007,tsalmantza2009} who developed a galaxy library for the
Gaia mission and explored classification problems therein, \cite{gao} who used
them to search for quasars in SDSS data, and \cite{huertas} who used them for
morphological galaxy classification. \cite{bailerjones2008} discussed SVM
classification of astrophysical sources in the context of unbalanced samples.

We proceed by taking the sample of \citeauthor{xue} and obtaining the
up-to-date photometry for it from SDSS DR7. We then investigate the ability of
each of the three techniques to recover the BHB stars from the Xue sample, and
the various options that are available to optimize them. Finally, we take a
new sample of DR7 photometry, for sources without spectra, and apply our
models to this sample to recover samples of probable BHB stars. We use a
selection of globular clusters of known distance to test the BHB
classifications and photometric parallaxes.

\section{Data}
\label{sect:data}

The latest publicly available SDSS data release, DR7, covers approximately
8400 square degrees, with images in the five SDSS bands: $u,g,r,i,z$.  Spectra
are available for a subset of the detected objects based on various selection
criteria.

The study of \citeauthor{xue} used  a sample of SDSS DR6 data selected to
  lie inside the colour box suggested by \cite{yanny} ($0.8 < u-g < 1.6$,
  $-0.5 < g-r < 0.0$). We have recovered the sources used by \citeauthor{xue} in
  the DR7 release by matching the SDSS MJD, plateId and fiberId fields.  We
obtained the PSF magnitudes, estimated extinction, and the parameters (Teff,
logg, and [Fe/H]) as determined by the SDSS pipeline. The dereddened
magnitudes were obtained from the model magnitudes during the pipeline
processing by applying extinction corrections derived from the map of
\cite{schlegel}. We recover the extinction from the model magnitudes 
and apply it to the PSF
magnitudes. The DR7 photometry is generally consistent with the DR6 photometry
given by \citeauthor{xue} to within hundredths of a magnitude, but there
are a number of sources with more divergent values. We rejected the most
discrepant of these by introducing a colour cut 0.1 magnitudes outside of
the colour cut of \citeauthor{yanny}. This cut excluded mostly contaminant
stars. The \citeauthor{xue} sample contained 10224 objects, of which 2558 were
identified by them as BHB stars.  After rejecting sources with discrepant
photometry, 9929 objects remained, of which 2536 were identified by
\citeauthor{xue} as BHB stars.

\begin{figure}[htbp!]
\begin{center}
\includegraphics[height=8.cm,angle=270.]{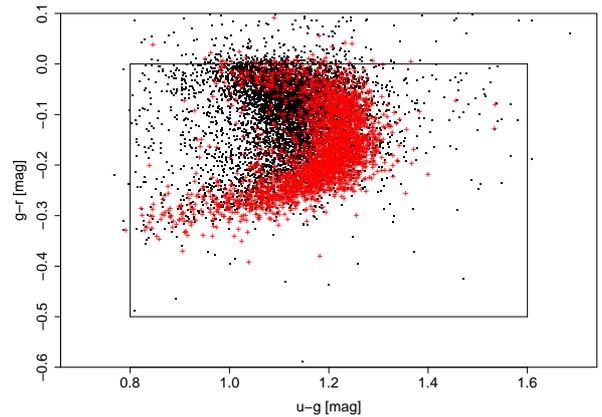}
\end{center}
\caption{The sample of \citet{xue} 
in the $u-g,g-r$ plane, with DR7 data. BHB stars (according
to \citeauthor{xue}) are shown as red crosses, non-BHB stars as black points. 
The box shows the colour cut used
by \citet{xue} and \cite{yanny}. 
The outer boundary extends 0.1 magnitudes beyond this. Sources 
outside the plot region were discarded.
\label{ugr_ccd}}
\end{figure}

We also cross matched against a list of 1172 objects from the paper of
\citeauthor{sirko}.  All these objects were identified by \citeauthor{sirko}
as BHB stars. Since \citeauthor{sirko} did not provide SDSS identities in
their table, we cross matched first with our SDSS data on the basis of RA and
Dec and $g$ magnitude. After cross matching, 1101 of the sources were
identified in the SDSS DR7 data. Of these, 4 had no identified counterpart
amongst the \citeauthor{xue} objects. Figure~\ref{ugr_ccd} shows the
colour-colour diagram of our sample.

\section{General Approach}
\label{genapproach}

All our classification methods are supervised, meaning that they require
samples of data for objects of known type in order to train a model, which
can then be applied to new data. Various parameters must be set to
optimize the classification, and in the end the reliability of the methods
relative to each other and in absolute terms has to be determined in some way.
For these reasons, we need a sample of test objects of known type on which we
can run our trained models. 
\begin{figure*}[htbp!]
\begin{center}
\vbox{
\hbox{
\hbox{
\includegraphics[height=4.5cm,angle=270.]{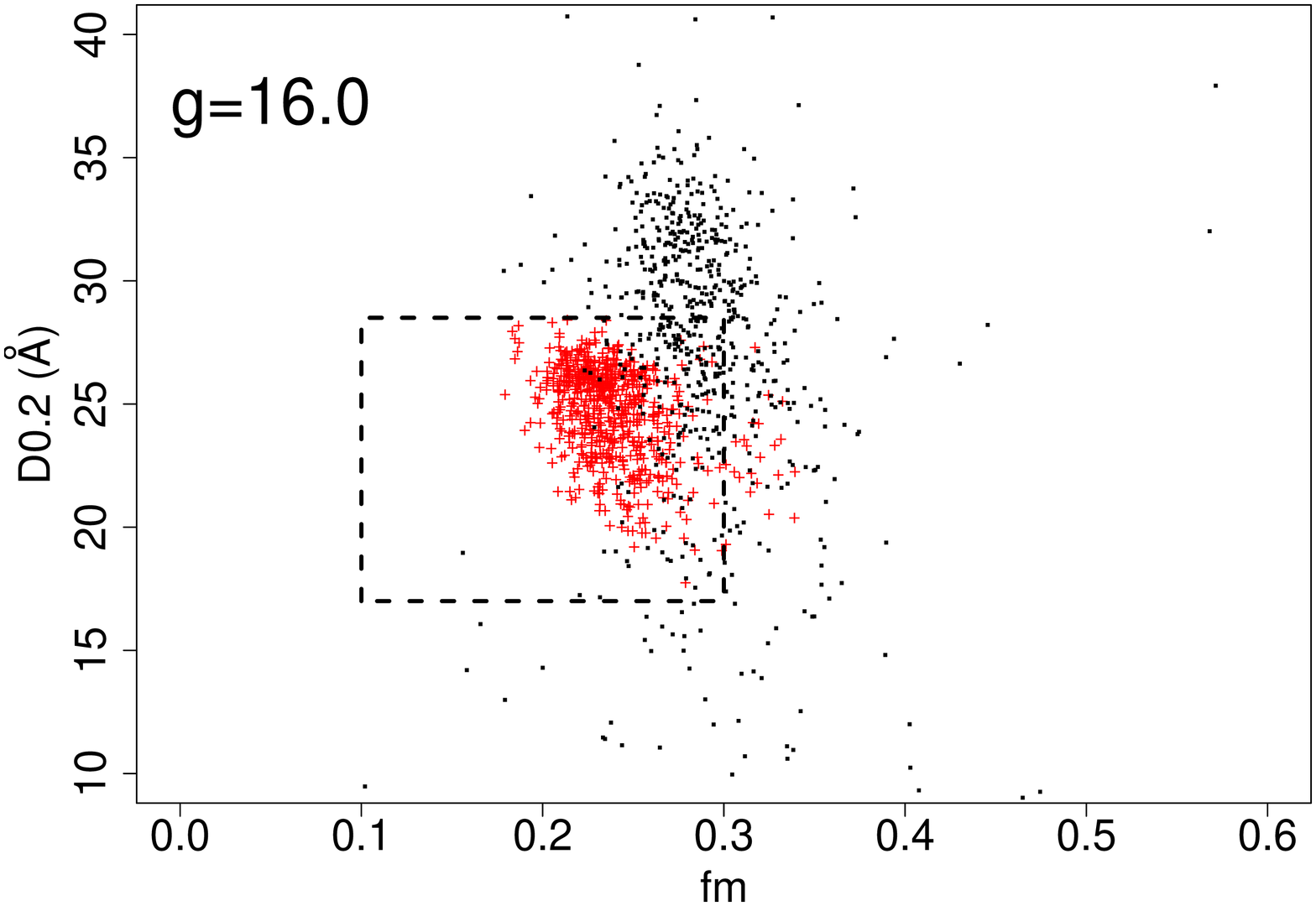}
\includegraphics[height=4.5cm,angle=270.]{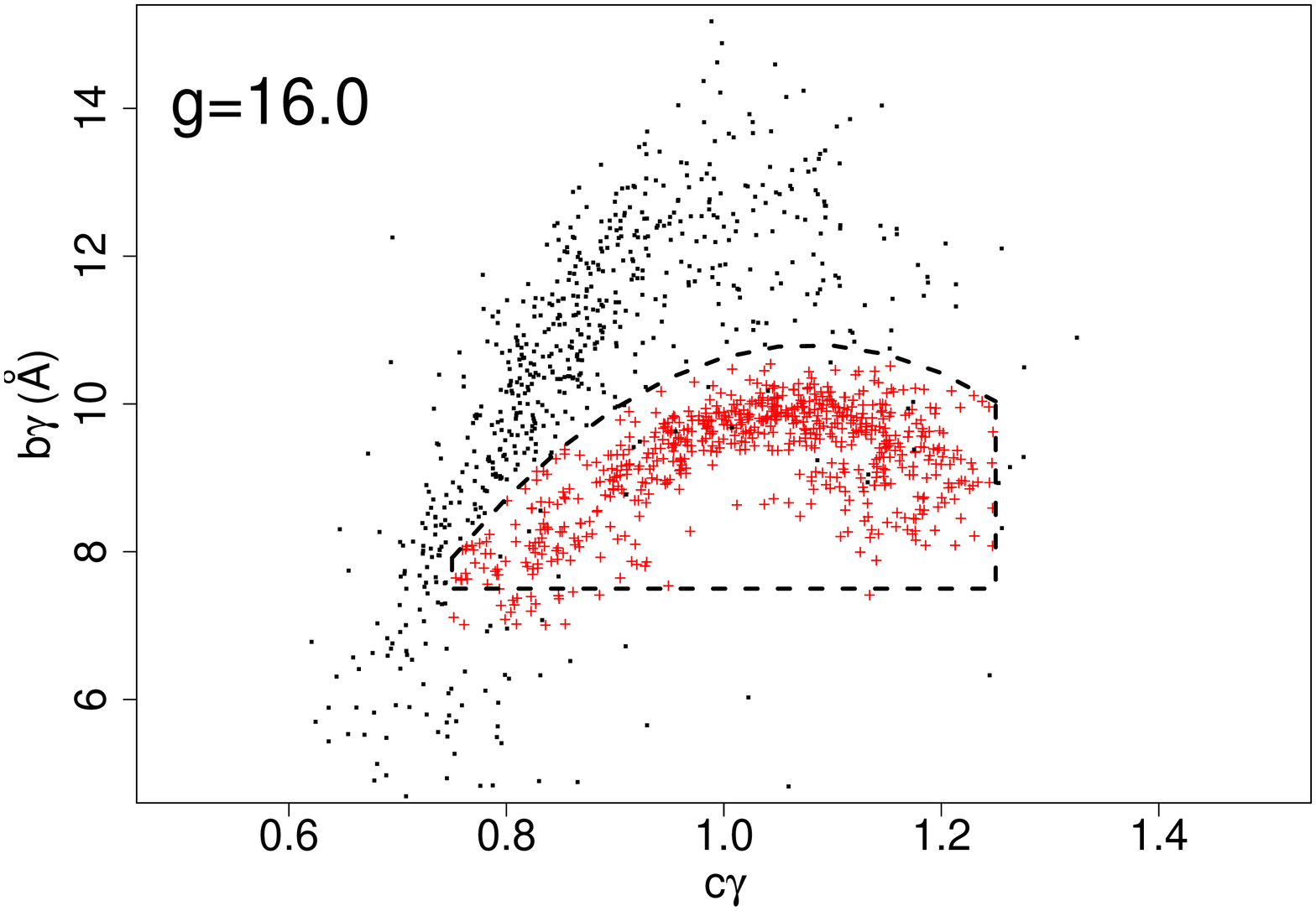}
}
\hbox{
\includegraphics[height=4.5cm,angle=270.]{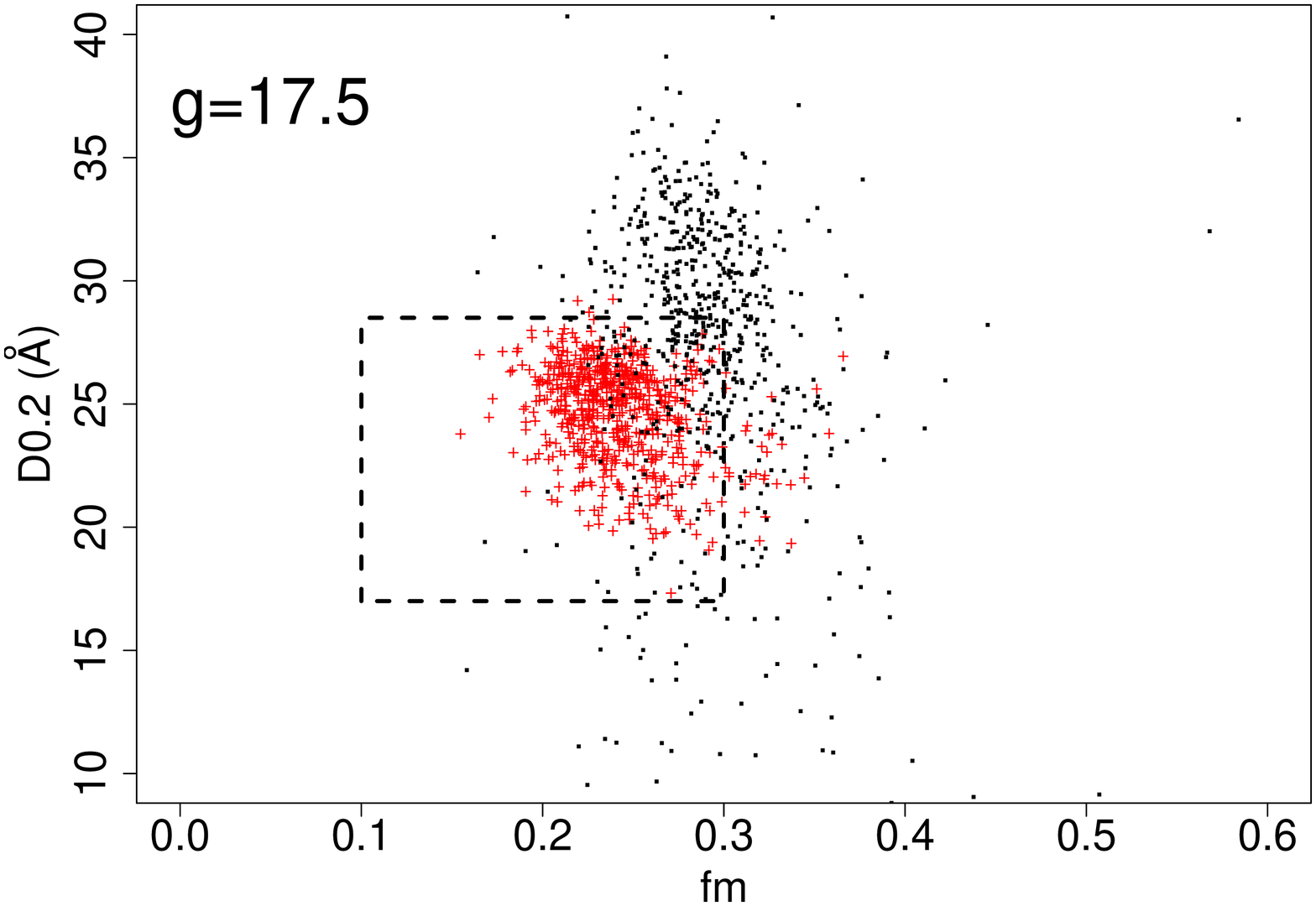}
\includegraphics[height=4.5cm,angle=270.]{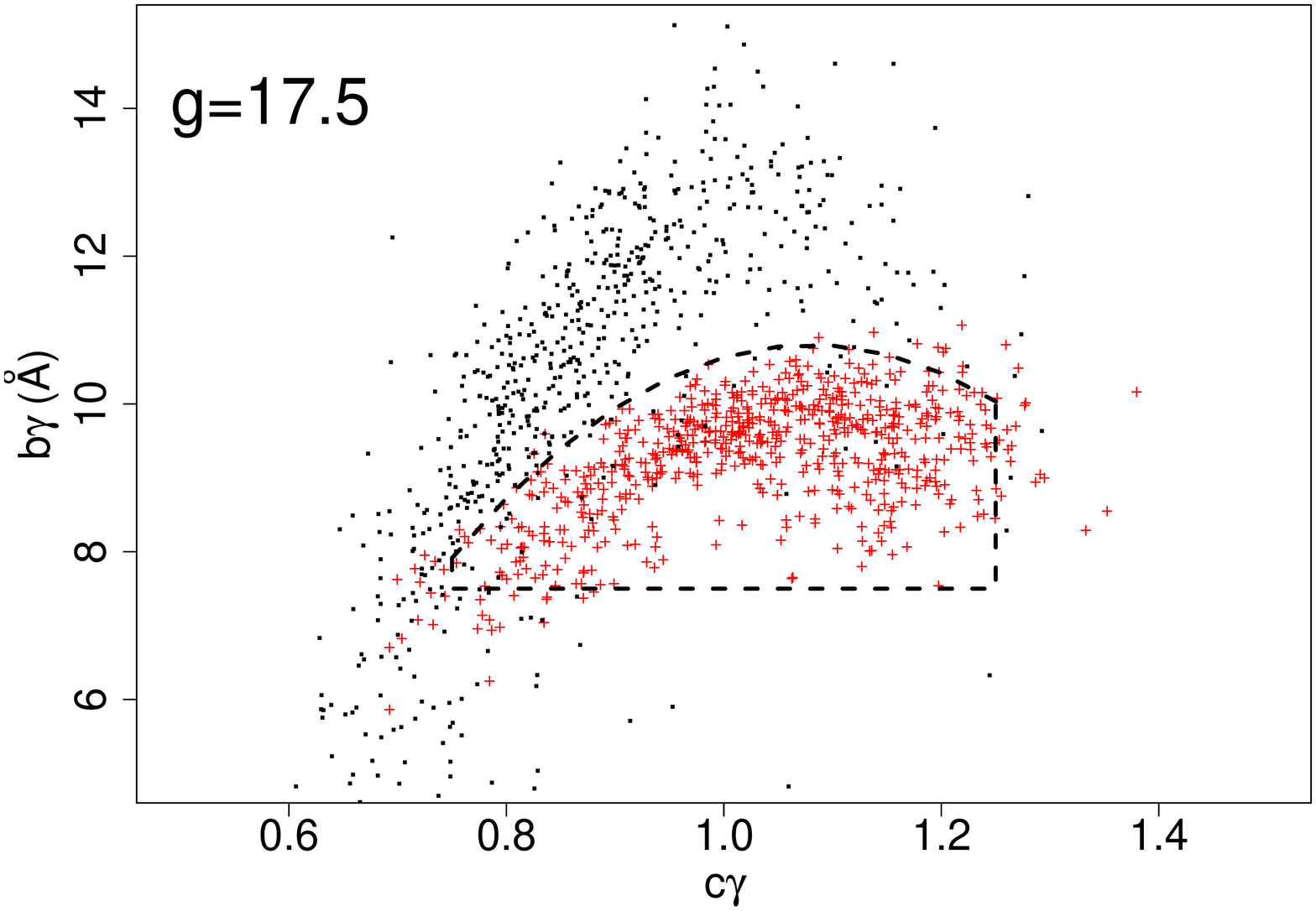}
}
}
\hbox{
\hbox{
\includegraphics[height=4.5cm,angle=270.]{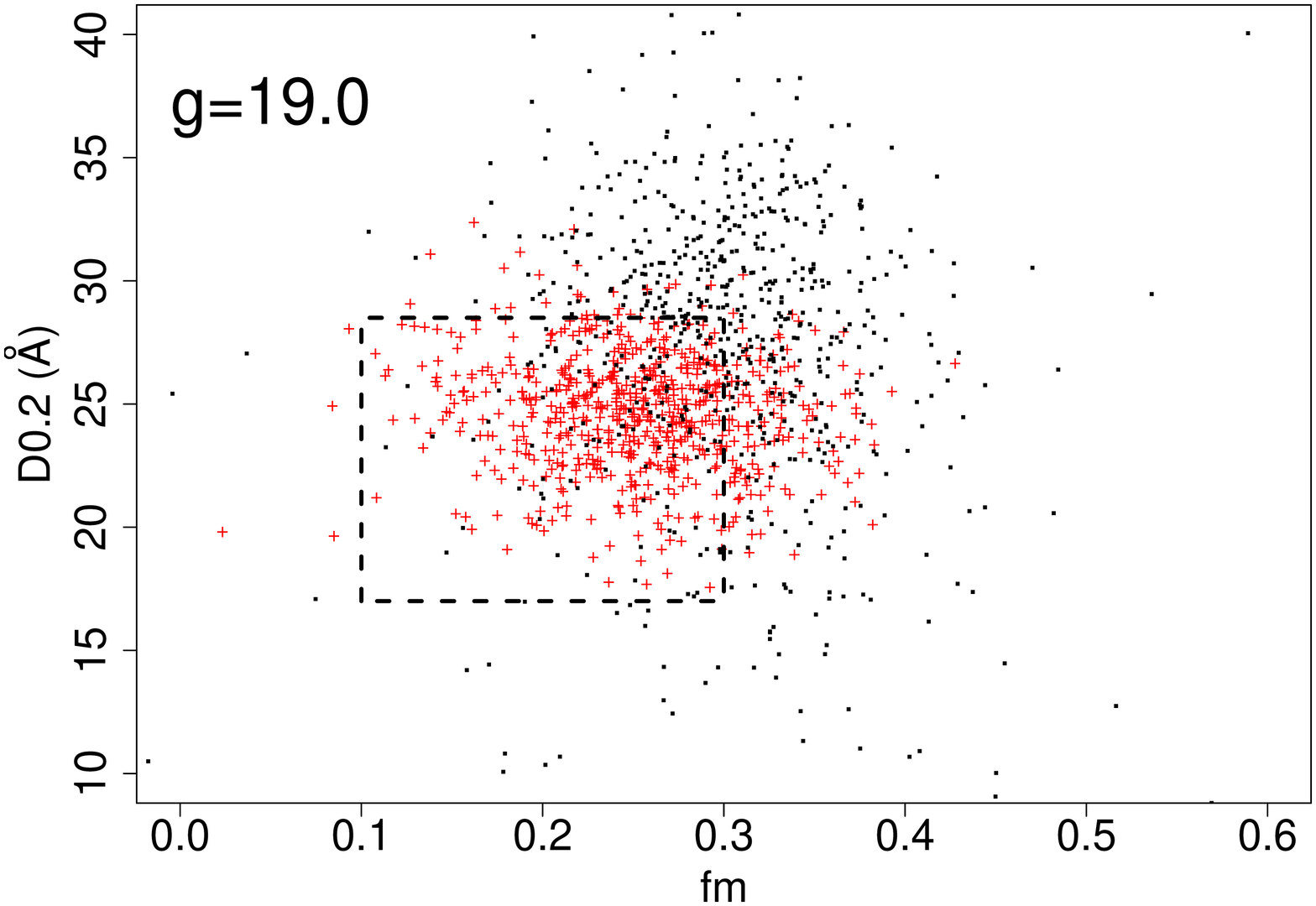}
\includegraphics[height=4.5cm,angle=270.]{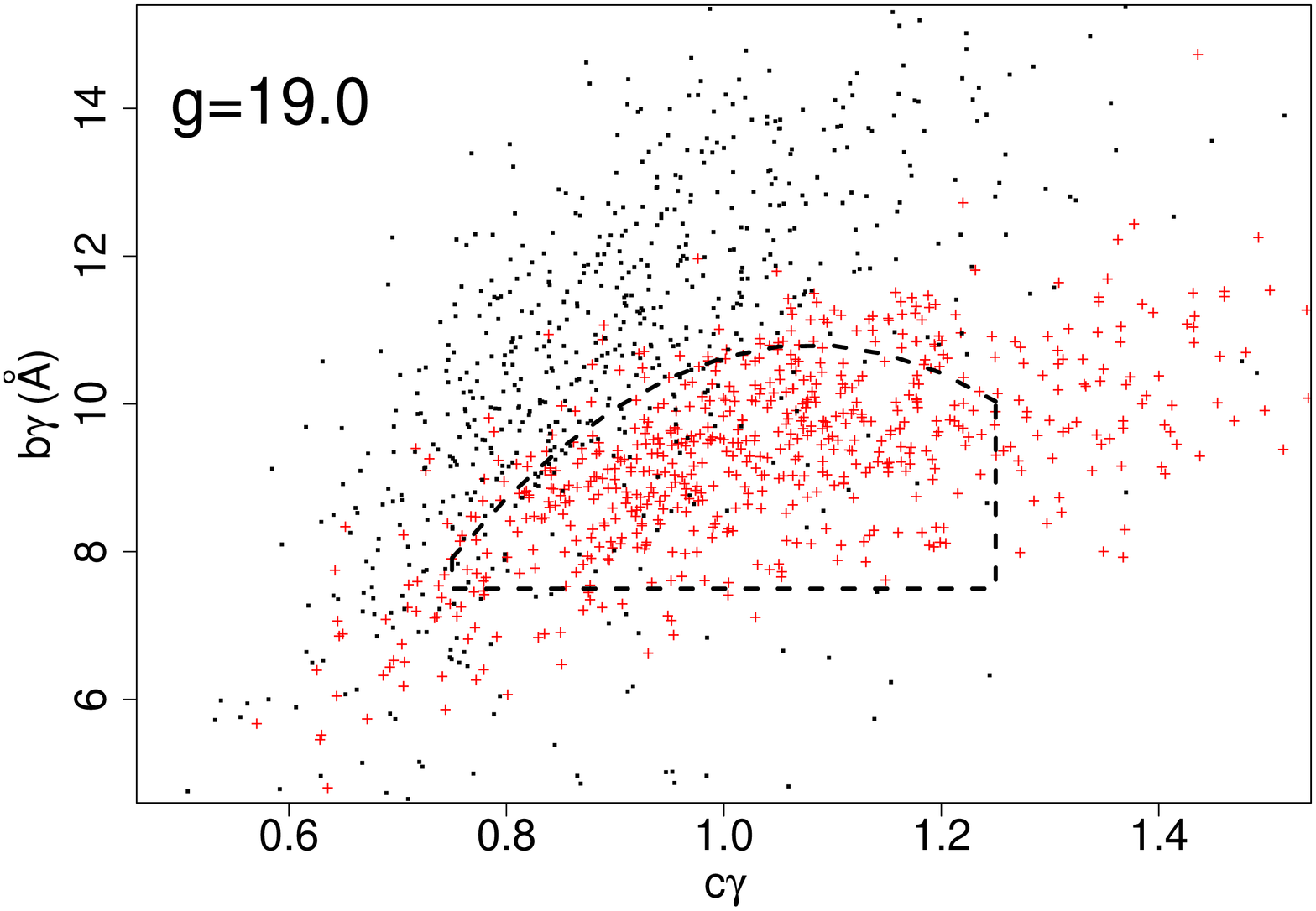}
}
\hbox{
\includegraphics[height=4.5cm,angle=270.]{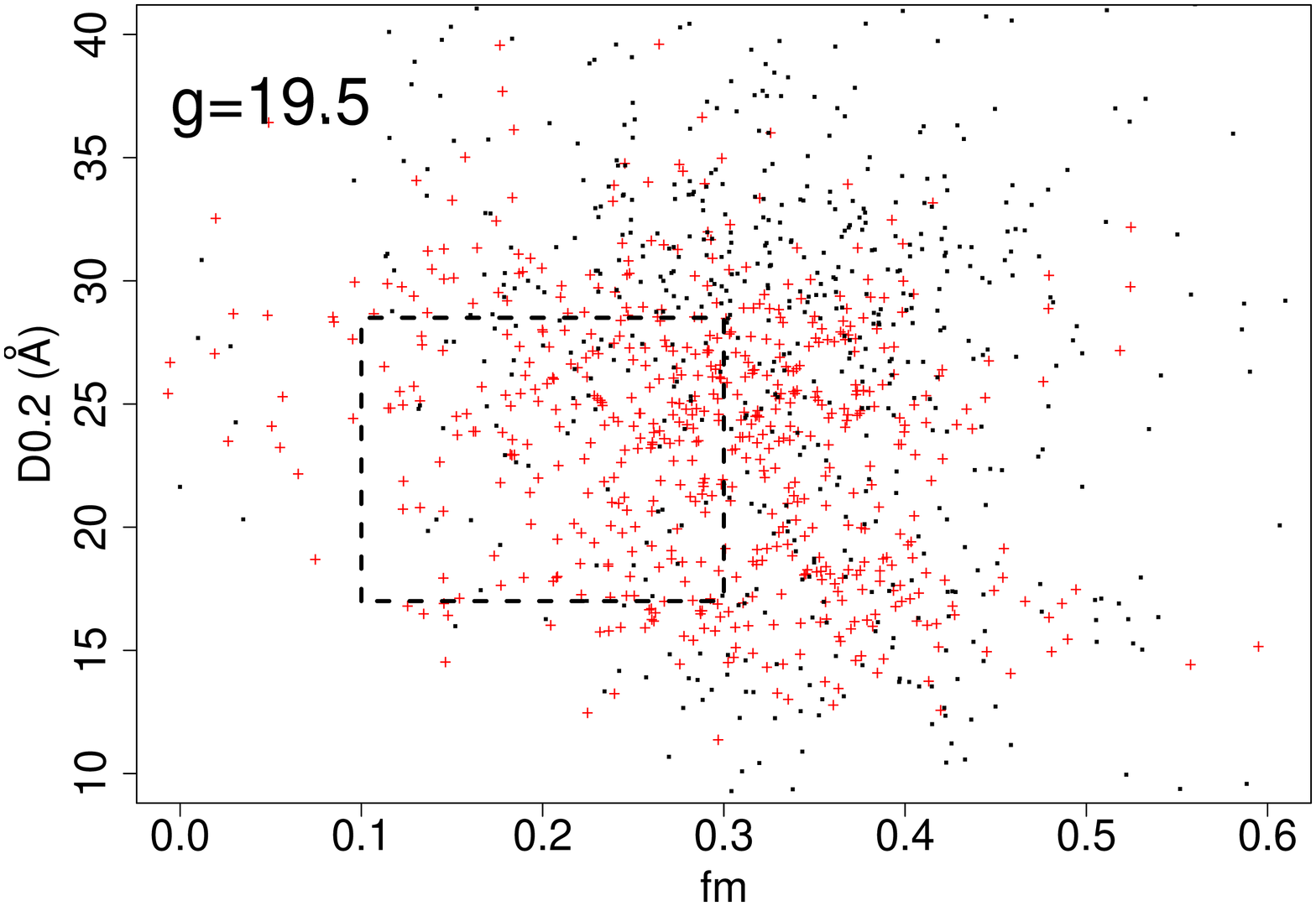}
\includegraphics[height=4.5cm,angle=270.]{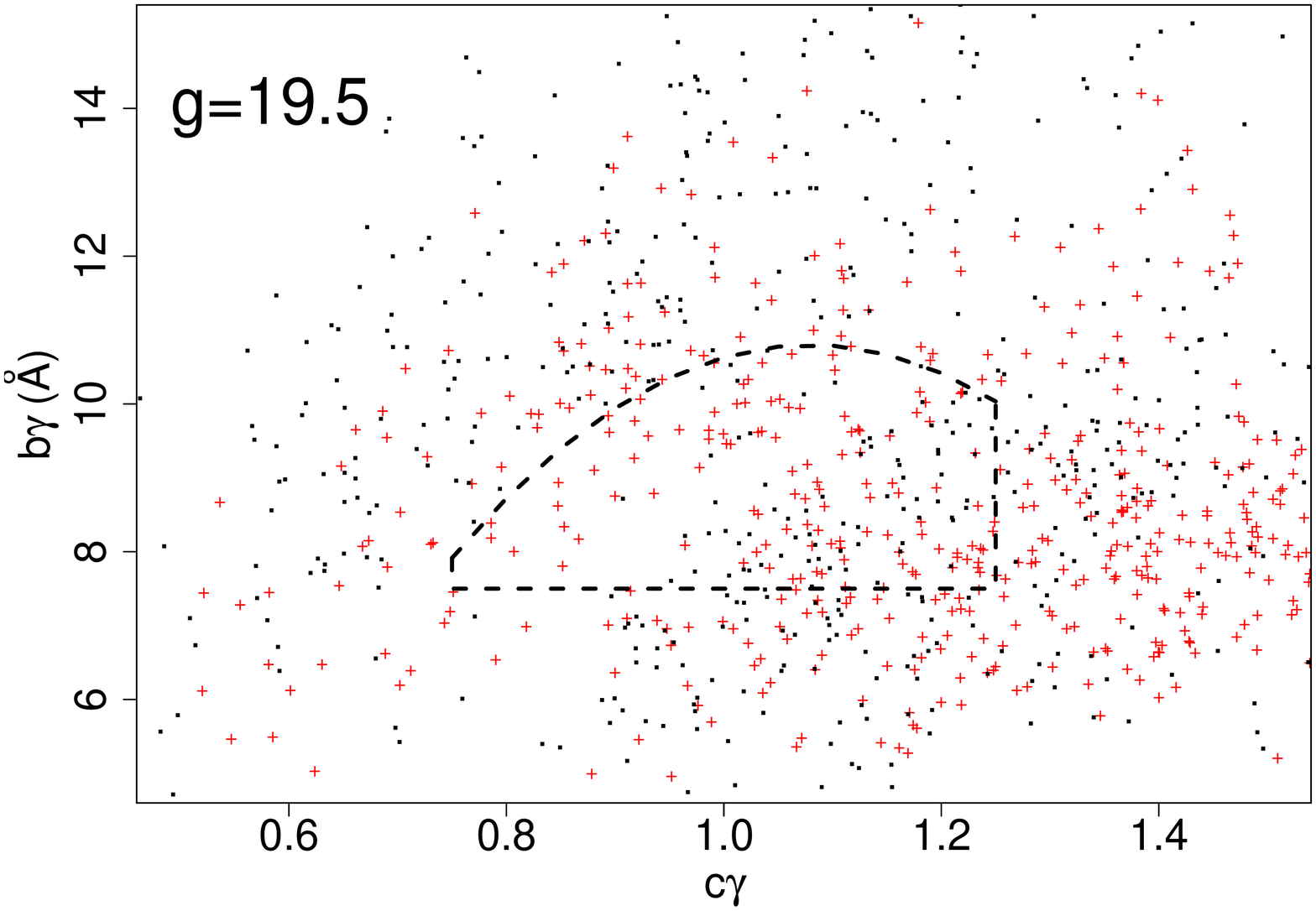}
}
}
}
\end{center}
\caption{Simulation of the effect of noise on \citeauthor{xue}'s 
classification. The four spectral 
line parameters used in the classification are 
fm, D0.2, $c_{\gamma}$, and $b_{\gamma}$ (see \cite{xue} for details). 
Four magnitudes are illustrated. for each of these, we plot 
D0.2 versus fm and $c_{\gamma}$ versus $b_{\gamma}$. 
The selection boxes are shown in each panel. The progressive
loss of BHB stars (red crosses) from the selection box is clear. Non-BHB 
stars (black dots) are not scattered into the boxes at the same rate. 
\label{xueReanalysisFig}}
\end{figure*}
Our sample contains 2536 BHB
stars as identified by \citeauthor{xue}  
Our standard procedure was to randomly
split the BHB sources into roughly equal training and testing sets, and then
randomly select equal numbers of non-BHB sources (designated ``other'') to
include in the training and testing samples. We can investigate the 
statistical properties of the results by bootstrapping.

\subsection{Data dimensionality and feature selection}

The PSF photometry was corrected 
for the expected extinction determined by the SDSS pipeline. 
There are four colours available.  The $u-g$
and $g-r$ colours are the most important for BHBs. The others show little or
no real information when examined by eye.  They make some difference (for the
better) for the kNN and SVM methods, but tend to degrade the KDE. It is
possible that the improvement seen for kNN and SVM by including the other
colours is mostly due to excluding faint sources due to their large scatter in
all bands.

\subsection{Comparing methods: completeness and contamination}

The {\em completeness} is defined as the number of correctly classified
sources of a particular class, divided by the number of available sources of
that class, i.e. it is the fraction of test sources of a particular class that
are correctly classified,
\begin{equation}
\mathrm{completeness}_j = \frac{n_{i=j,j}}{N_i},
\label{complEqn}
\end{equation}
where $n_{i,j}$ is the number of objects of true class $i$ classified as 
output class $j$ and $N_i$ is the total number of input sources of class $i$. 
Input sources can be lost
from the output class due to misclassification into another class, or by
remaining unclassified due to an insufficiently high classification
confidence. The {\em contamination} of the output sample is defined as the
number of falsely classified sources of that class divided by the number of
sources classified into that class, whether correctly or incorrectly, 
\begin{equation}
\mathrm{contamination}_j = \frac{\sum_{i \ne j} n_{i,j}}{\sum_{i}n_{i,j}},
\label{contamEqn}
\end{equation}

In our particular case, one class, the set of non-BHB stars, is really a mixed
class of contaminants comprising blue stragglers and main sequence stars, that
we are interested in removing in order to obtain a clean sample of BHB
stars. Therefore, we are interested in the completeness and contamination of
the BHB sample and not primarily in the completeness or contamination of the
``other'' class.

\subsection{Reliability of the training and testing sets}
\label{xueReanalysisSect}

Since our method is based on the results of \citeauthor{xue}, it is worthwhile
investigating how reliable these are, particularly at the faint end. To this
end, we selected 1381 spectra from \citeauthor{xue}'s original
data, having $14.5<g<15.5$. Roughly half of these (655) were BHB stars. 
We then added artificial noise to degrade them to the same
signal-to-noise ratio as fainter spectra. We constructed in this way 
eight artificial samples in half magnitude steps from $g$=16.0 to $g$=19.5. 
We then reanalysed these degraded spectra with the technique of 
\citeauthor{xue} and reclassified them. We then compare the performance 
at faint magnitudes with the original performance. 

The classification of \citeauthor{xue} is based on recovering four different
characteristic parameters from the absorption lines. These are D0.2, the width
at 20\% below the continuum of the Balmer line, fm, the flux relative to the
continuum at the line core, and $c_{\gamma}$ and $b_{\gamma}$, which are
parameters from a S\'ersic fit to the line shape \citep{sersic}.  
BHB stars are identified as lying
within selection boxes in the feature space formed by the line
parameters. This selection is illustrated for our degraded data in
Fig.~\ref{xueReanalysisFig}, which shows for four example magnitudes the
D0.2 versus fm and $c_{\gamma}$ versus $b_{\gamma}$ values, and the selection
boxes.  It is clear from the figure that, as the noise increases, true BHB
stars scatter outside one or the other selection box and are lost, decreasing
the completeness. Some sources from outside the selection boxes scatter into
the box, but since the box covers a small fraction of the data space, and
since a contaminant has to scatter into both boxes to be misclassified as a
BHB, the increase in the absolute number of contaminating sources is
modest. The contamination, as defined above, will still 
increase because the
number of true positives, in the denominator of Equation~\ref{contamEqn}, 
is decreasing.
Figure~\ref{reanalysisResults} shows the resulting ratio of objects classified
(rightly or wrongly) as BHB stars to total stars as a function of $g$.   
This is discussed in terms of the effect on the prior
probability in the following section. For now, we note that this effect kicks
in strongly for sources fainter than $g$=19.0, and that it will degrade the
quality of training sets used to define models and also of any testing set
used to assess them.

\subsection{Priors}
\label{sect:priors_general}

The classifiers we use are trained on mixed samples of BHB and non-BHB stars
with a range of properties. We implicitly assume that 
the classifier takes account of the likely distribution of the 
population of objects to be clasified, and if it does
not, we need to correct the classifier output probabilities 
using appropriate priors. 
\begin{figure}[hbp!]
\begin{center}
\includegraphics[height=8.cm,angle=270.]{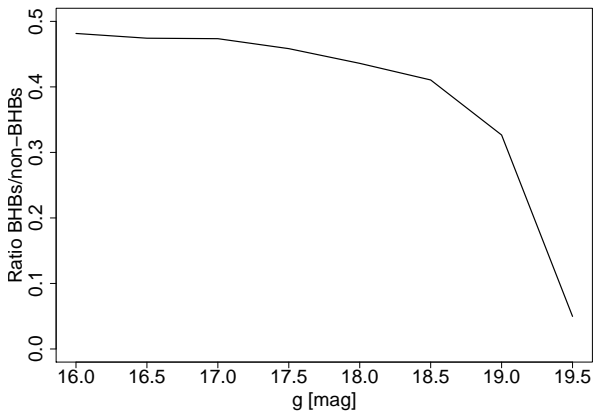}
\end{center}
\caption{The effect of increasing noise on the spectroscopic classification
of \citeauthor{xue}, as illustrated in Fig.~\ref{xueReanalysisFig}. 
The line shows the ratio of the number of {\em output} BHB stars, 
that is, the sources classified as BHB stars, regardless of 
whether they really are 
or not, to total stars.
\label{reanalysisResults}}
\end{figure}

The simplest prior to be accounted for is the true class fraction.  To train
our models, we use equal numbers of BHB and non-BHB stars, whereas the true
class fractions are not equal (the fraction of BHB stars in the sample of
\citeauthor{xue} is approximately 0.26 for all sources). For the kNN and KDE
methods we could directly include the class fractions in the training sets.
For the SVM we could also include proportional fractions of classes, but the
actual effect of this on the classifier is complex and not well understood.
We choose to always use equal class fractions and use a prior to adjust
the classifier output. We refer to this type of class fraction prior as a 'simple prior'  hereafter.

\begin{figure}[htbp!]
\begin{center}
\includegraphics[height=8.cm,angle=270.]{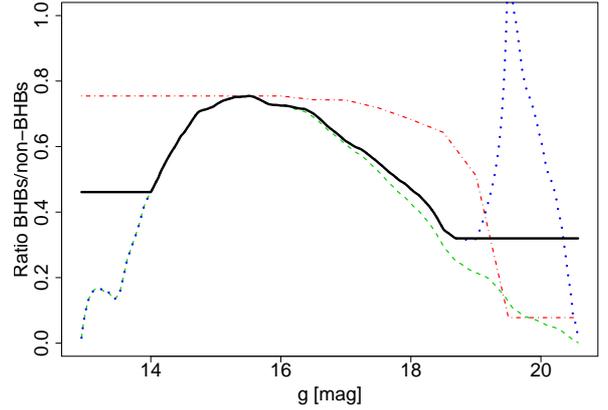}
\end{center}
\caption{The dashed green line shows the ratio of the density function 
of BHB stars to the density function of all stars. 
The dot-dashed red line shows the ratio of BHB stars to non-BHB 
stars from the classification described in Sect.~\ref{xueReanalysisSect}. 
This is the same curve plotted in Fig.~\ref{reanalysisResults} but 
it has been renormalized so that the peak is at the same level 
as the peak of the ratio of density functions - i.e. so that it doesn't 
include the simple prior. The dotted blue curve 
is the result of correcting the basic BHB fraction (dashed green line)
for the expected 
change in the ratio due to noisy spectra. 
The regions at either end are replaced with a constant value, and 
the prior actually adopted is 
plotted as the thick solid black line.
\label{magpriors}}
\end{figure}

We can also adjust for prior probabilities as a function of other 
parameters that are not accounted for by the classifier itself. We consider 
$g$ magnitude and $b$, the Galactic latitude. 
In Fig.~\ref{magpriors} we show the ratio of the density functions of BHB
stars and all stars as functions of $g$ in the sample of \citeauthor{xue}
(dashed green line). Because these density functions were individually
normalized, the resulting ratio is the relative fraction 
of BHB stars, rather than the absolute fraction - i.e. 
it is as if the class fractions were equal.  

Also shown in this plot is the ratio of BHB stars to all sources at each
magnitude as estimated from the experiment described above in
Sect.~\ref{xueReanalysisSect} (red dash-dotted line). This curve, which
  is a renormalized version of the curve shown in
  Fig.~\ref{reanalysisResults}, shows what we would expect to see if the
fraction of BHB stars in the true population was constant with $g$, but the
observed ratio was altered by sources being lost due to increasing noise at
high magnitudes. We can correct the measured ratio of BHBs to all sources for
this effect. This correction mitigates the falloff of the ratio at the faint
end. The corrected curve is shown as the dotted blue line.  
We use this as
the basis of the magnitude dependent prior, but the correction causes a spike
at the faint end that is probably due to small numbers of sources and is
obviously not desirable in the prior. For this reason we have truncated the
function before it turns up and adopted a plateau for the high magnitude end.
Similarly, we adopt a plateau at the bright end, where the fraction may be
strongly affected by the SDSS spectrum selection function 
(there is a cutoff
at $g<14$ for the Legacy spectra, \citep{dr4paper}).  The adopted prior as a
function of $g$ is plotted as the solid black line in Fig.~\ref{magpriors}.

\begin{figure}[htbp!]
\begin{center}
\includegraphics[height=8.cm,angle=270.]{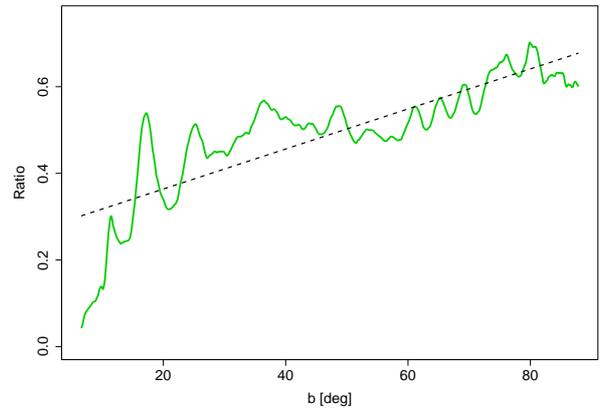}
\end{center}
\caption{The solid green line shows the ratio of BHB star density to 
the sum of BHB density and other stars 
density as a function of galactic latitude $b$.
The dashed black line is a linear fit used to build the 2-dimensional 
prior as described in the text.  
\label{latitudes}}
\end{figure}

Figure~\ref{latitudes} shows the ratio of density functions for BHB and
non-BHB sources as a function of absolute Galactic latitude. 
This ratio shows a relatively smooth trend, with
quite a lot of structure superimposed. We model it with a straight line
fit and adopt this fit as the relative prior in latitude. 

If these priors are independent of one another, we can apply them 
in sequence to the output posterior probability of the classifier 
using
\begin{equation}
P(C|D_1,D_2,....,D_N) = \Pi^N_{n=1} \frac{P(C|D_n)}{P(C)^{N-1}},
\end{equation}
where $P(C|D_n)$ is the probability of class membership given some piece of
information, $D_n$. This formula is discussed in depth in \cite{caljTN53}.
The issue of class fraction priors and its influence on classifier training is
discussed in \cite{bailerjones2008}. The correlation of $g$ and $b$ is low,
with a Pearson coefficient of -0.0017. The assumption of independence
therefore holds well. 

The ratio of BHB stars to all stars for all data is 0.26, however this
includes regions where the selection function for SDSS spectra has a large
effect (there is a cutoff at $g=14$ for SDSS Legacy spectra and $g$,$r$ or
$i$=15 for SEGUE, and at $g>19$ the reliability of the \citeauthor{xue}
classification method becomes difficult to assess).  The ratio of BHB stars to
all stars in the interval $14<g<19$ is 0.32. We use this latter fraction as
the class fraction.

For the analysis of the different classifiers, we consider two different
priors, a simple ratio (equal to 0.32) that represents the fraction of BHB
stars to all stars over the sample in the interval $14<g<19$, and the
combination of this with the priors as functions of $g$ or $b$ as discussed
above. We refer to the first prior as a 'simple prior' and to the second as a
'2d prior', because it is a function of the two variables $g$ and $b$.

\subsection{Priors and performance measures}

The issue 
of the class fractions enters the analysis in two distinct ways, and
it is worth discussing this issue explicitly because it can easily lead to
confusion.

Firstly, the class fractions are the single most important contribution
to the prior probability used to obtain posterior probabilities for each
object. This issue is reasonably clear.

Secondly, as well as adjusting the classifier probabilities with the
prior, when testing the classifiers, we 
{\bf also} have to take account of the uneven
expected class fractions in the measured contamination (and in any other
quantity where they would be important -- the completeness is not affected, as can be seen from Equation~\ref{complEqn}).
We can do this by either using a test set that reflects the true expected
fractions (which would be 
possible in our case because the classes are not extremely
unbalanced), or by correcting the contamination for the difference between the
input test set fractions and the expected true fractions.  Since the
population composition as a function of $g$ and $b$ is already present in the
test set, no correction should be made to the test output to correct for the
relative fractions of BHB stars as a function of these quantities.

Note that this reweighting of the contaminants in the output sample has to be
carried out anyway if the expected class fractions are different to the
fractions in the test set, whether or not we also apply the prior to the
classifier probabilities. All the contaminations presented in this paper,
except that in Table~\ref{yannyResults}, are based on test sets with equal
class fractions, and are corrected to the expected class fractions after the
classification using the estimated fraction of 32\% BHB stars.

The issue of priors in the context of a classification problem with a highly
unbalanced data set was addressed in some depth by \cite{bailerjones2008}. In
particular, Sect.~2.5.1 of this paper discusses the issue of correction of
the contamination in more depth than is possible here. There the specific
problem was identifying quasars amongst stellar samples, which is an extremely
unbalanced problem. The issue with BHB stars is less severe.

\section{Comparative performance of machine learning techniques}

\subsection{Colour box and direct decision boundary}

\cite{yanny} derived a decision boundary in $u-g$ $g-r$ colour space to
distinguish low gravity BHB giants from contaminating MS and BS stars in the 
colour box. Their decision boundary consists of three straight line segments
and is shown in their Fig. 10. Our estimates of the 
gradients and intercepts of the 
line segments are shown in Table~\ref{yannyDecision}. 
Sources are BHB stars if they 
have $u-g < m(g-r)+c$, with values of $m$ and $c$
taken from the Table.
\begin{table}
\caption{Coefficients for decision boundary in $u-g$, $g-r$. 
\label{yannyDecision}
}
\begin{tabular}{ccc}
Segment                  & $m$      & $c$ \\
 $ -0.3  < g-r <= -0.25$ & 2.4      & 1.62 \\
 $ -0.25 < g-r <= -0.15$ & 1.6      & 1.42 \\
 $ -0.15 < g-r <= 0.$    & -0.533   & 1.1  \\
\end{tabular}
\end{table}

We classified our test set with this boundary, and obtained the results
summarized in Table~\ref{yannyResults}. All sources with $g<19$
were classified, since no training set is needed. The test set class fractions
are not artificially balanced, so no prior has been applied.
Sources lying outside the ranges of
Table~\ref{yannyDecision} remain unclassified. 
The results are presented first in the form of a confusion
matrix. Each row of the matrix corresponds to a particular true class, either
BHB or other. The rows are labeled in capitals to indicate that this is the
true class of the object. The columns list the output classifications. The
leading diagonal of the matrix therefore shows the true classifications. The
off diagonal elements indicate misclassifications, and it is possible to see
which classes are particularly confused with one another.  The confusion
matrix is presented twice, once with the absolute numbers of objects in each
classification bin, and once with the classifications expressed as
percentages of the total number of input objects of that class. The rows of
this matrix therefore sum to one. We also present the completeness and
contamination obtained with this method.

\begin{table}
\caption{Results of classification with decision boundary.  
\label{yannyResults}
}
\begin{tabular}{lll}
\multicolumn{3}{c}{Absolute} \\
       & bhb     &    other     \\
BHB    & 1963    &    486       \\
OTHER  & 2560    &   3321       \\  
Unclassified & 436          \\
 \multicolumn{3}{c}{Percentage} \\
BHB    & 80.16   &   19.84     \\
OTHER  & 43.53   &   56.47      \\
       &  Completeness    & Contamination \\ 
BHB    &  0.802  &   0.566     \\
\end{tabular}
\end{table}

\subsection{k-Nearest Neighbours}

Nearest neighbour techniques are probably the simplest and most intuitively
obvious method for supervised classification. For a given new object, we
select the $k$ nearest training points in the data space and assign a class
based on the classes of the neighbours. For $k>1$ we could choose to select a
simple majority of objects, or we could impose a higher threshold in an
attempt to improve the purity of one or both of the output classes (i.e. BHB
stars or other).  Introducing a threshold implies that we must be prepared to
tolerate non-classifications.
\begin{figure}[htbp!]
\begin{center}
\includegraphics[height=9.cm,angle=270.]{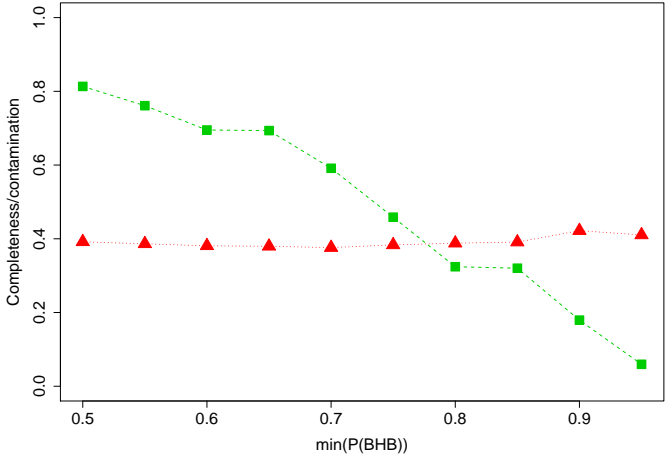}
\end{center}
\caption{The effect on the completeness (green squares) 
and contamination (red triangles) 
of varying the minimum probability (including the effect of the 2d prior)
  required for a positive classification in the kNN method. The completeness
  (necessarily) falls as the threshold is increased.  
  \label{knn_thresholds_plot}}
\end{figure}

A probability can be estimated from the fraction of neighbours belonging to
each class, so for example if nine out of ten of the nearest neighbours are
BHB stars, we would estimate $P(BHB)=0.9$. 

We ran the kNN technique for various choices of $k$ and measured the output
sample completeness and contamination.  The classifier was run with ten
resamplings of the training and test sets for each $k$, from $k=1$ up to
$k=100$, and the classification was performed with simple majority voting.
The completeness was found to be approximately constant with increasing $k$,
but the contamination showed a shallow minimum at around $k=15$, which we
selected as the optimum value.

We also experimented by cutting the colours used from four down to two ($u-g$
and $g-r$). The result was a slight degrading of the results for all values of
$k$. We therefore use the kNN technique with the four dereddened colours.

We next investigated the effect of varying the 
confidence threshold for classification and measuring the completeness and
contamination of the output BHB star sample. Increasing the threshold 
would be expected to lead to a loss of completeness, but 
also a lowering of the contamination.  The results of this for kNN 
are shown in Fig.~\ref{knn_thresholds_plot}. 
The completeness does indeed fall, but the contamination remains constant at around 0.4.

We performed ten classifications, with resampled training and testing sets,
with the kNN method to get a final estimate of performance. The value of $k$
and the probability threshold were left fixed ($k=15$, threshold=0.5).  The
results are shown in Table~\ref{knn_basic}. This table is divided into three
sections. In the top section, we present the results of applying the
classifier to the test data without applying any prior. This is equivalent to
assuming equal true class fractions. The second section presents the results
with the application of the so-called simple prior, with which we correct for
the effect of the class fractions only. The final section presents the results
with the application of the 2d prior, a function of $g$ and $l$. In each
section we present the results as confusion matrices of absolute
classifications and as percentages of the input true classes. Finally we
present the completeness and contamination. 
\begin{center}
\begin{table}
\caption{Results for kNN classification.
\label{knn_basic} }
\begin{tabular}{lll} 
With equal priors   & &       \\   
\multicolumn{3}{c}{Absolute} \\
        &  bhb            &  other  \\
BHB     &  1008           &  235    \\
OTHER   &   347           &  871   \\  
\multicolumn{3}{c}{Percentage} \\
BHB     &  81.09           &   18.91    \\
OTHER   &  28.49           &   71.51   \\  
        &                  &             \\
        &  Completeness    & Contamination \\ 
BHB        &    0.811         &  0.422   \\ 
        &                  &        \\ \hline
With simple prior & & \\
\multicolumn{3}{c}{Absolute} \\
        &  bhb           &  other \\
BHB     &  743          &   500   \\
OTHER   &  245           &  973  \\
\multicolumn{3}{c}{Percentage} \\
BHB     &   59.77        & 40.23 \\
OTHER   &   20.11        & 79.89  \\
        &                 &           \\
        &  Completeness   & Contamination \\ 
BHB     &  0.598         &  0.412 \\ 
        &                  &        \\ \hline
With 2d prior  & & \\
\multicolumn{3}{c}{Absolute} \\
        &  bhb           &  other \\
BHB     &   920          &   323   \\
OTHER   &   264          &   954  \\
\multicolumn{3}{c}{Percentage} \\
BHB     &  74.01          & 25.99 \\
OTHER   &  21.67          & 78.33  \\
        &                 &           \\
        &  Completeness   & Contamination \\ 
BHB     &  0.740         &  0.378 \\ 
\end{tabular}
\tablefoot{{\em Top section:} 
Confusion matrix showing the results of kNN classification (k=15) 
with threshold $P(BHB)>0.5$. This is the combined result of ten independent 
classifications with resampling of the training and testing sets.
The rows show the true class (according to \citeauthor{xue}), the columns 
show the classifier output class. When shown as percentages, the 
quantities in the rows should add to 100\%, but the quantities in the 
columns in general do not.  
The completeness and 
contamination are also shown - and the contamination is 
corrected for the class imbalance using the simple prior. 
{\em Middle section:}  The same confusion matrix, with output 
probabilities corrected for the simple prior.  
{\em Bottom section:} The results with output probabilities 
corrected for the prior probability as function of $g$ 
magnitude and galactic latitude. }
\end{table}
\end{center}

\begin{figure}[htbp!]
\begin{center}
\vbox{
\includegraphics[height=8.cm,angle=270.]{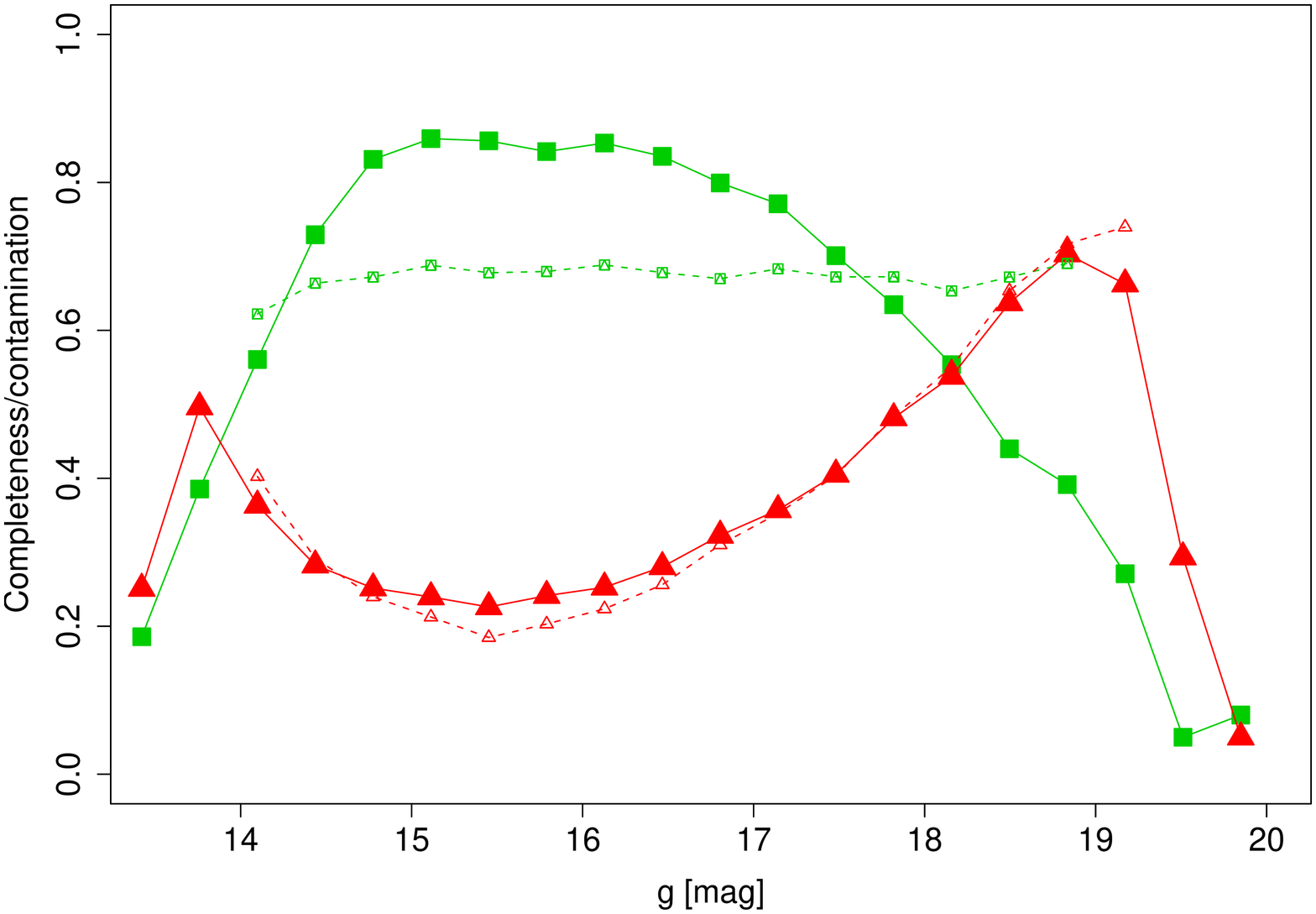}
\includegraphics[height=8.cm,angle=270.]{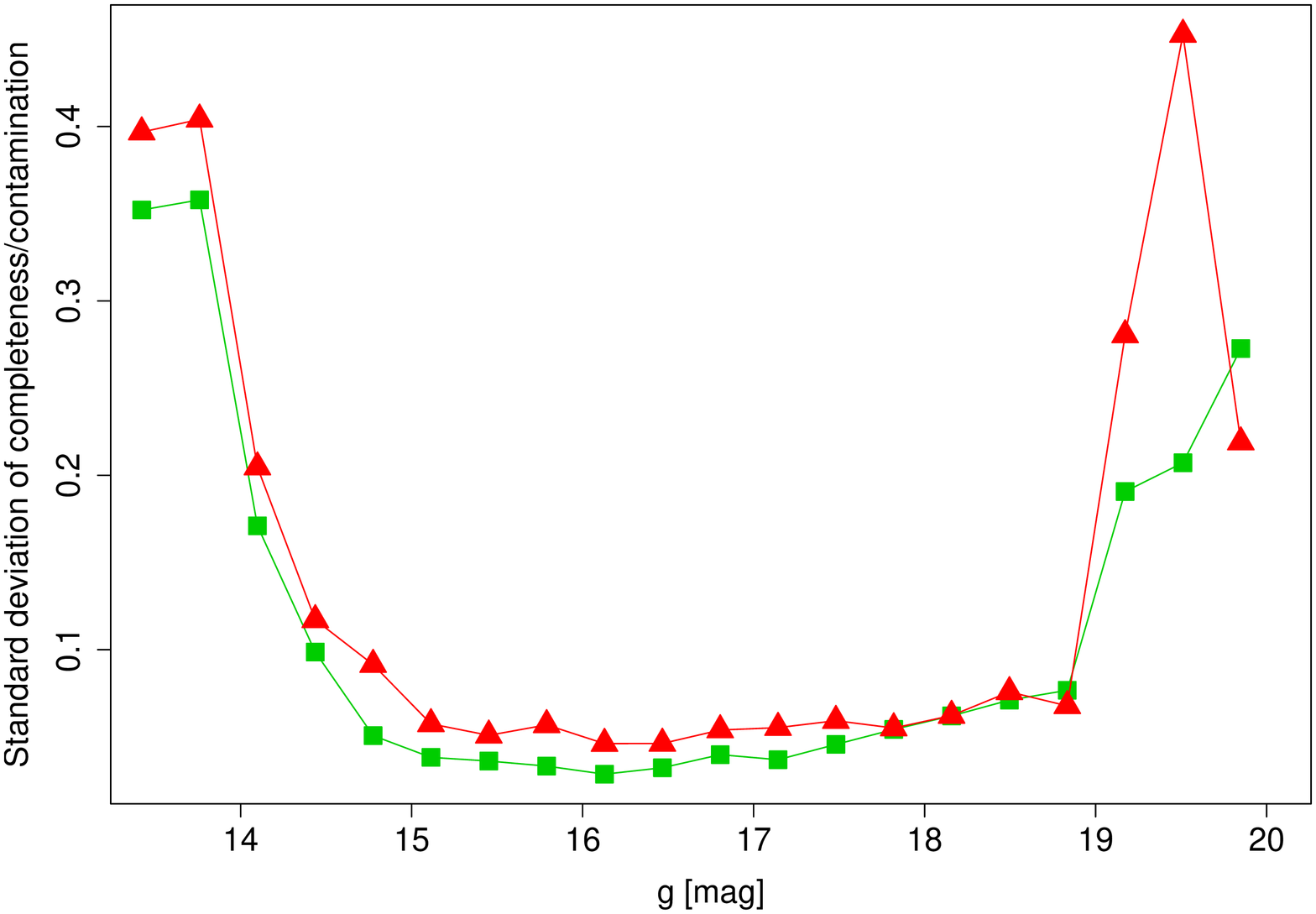}
}
\end{center}
\caption{{\em Top:} Completeness (green squares) and contamination 
(red triangles) 
for sources of different magnitudes classified with the kNN technique.
One hundred separate trials were averaged to produce this plot.
The filled symbols and solid lines show the results using the 2d prior.
The open symbols and dashed lines show the results using the simple 
prior only.  
{\em Bottom: } Standard deviations of completeness and contamination 
at each point.
\label{magdepcompcontamKnn}}
\end{figure}

Figure~\ref{magdepcompcontamKnn} shows the completeness and contamination for
test samples classified with the kNN method, with the results binned by
magnitude. The threshold for classification is always 0.5, and $k$=15. This is
the average of 100 separate trials. The lower plot shows the standard
deviation in each bin.

As expected, the classifier performance falls off for fainter magnitudes. 
Part of this effect may be due to the natural confusion in the test set between 
BHB stars and non-BHB stars, introduced by the noise in the method of
\cite{xue}. 

\subsection{Kernel density estimation for classification}
\label{sect:kde}

We next consider a kernel density estimation (KDE) approach to the
classification. The density estimate is a weighted mean of neighbours, the
weighting function being a kernel of choice.  See \cite{hastie} for a
 general discussion of the method, and see \cite{richards2,richards1} for
  examples of KDE used to identify quasars in SDSS data.

We use an Epanechnikov kernel, which is truncated and so
is less influenced by distant points. In practice, the
choice of kernel is usually less important than the bandwidth value. The
bandwidth was set independently for each dimension. The package {\em np} in
R\footnote{\url{http://www.r-project.org}} was used to implement the KDE
method \citep{np}, and also to determine the optimal value for the bandwidth,
using the method of \cite{liAndRacine}. This is based on leave one-out-cross
validation and involves minimizing the variance amongst trial density
functions constructed with different bandwidth values.

\begin{figure*}[htbp!]
\begin{center}
\hbox{
\includegraphics[height=9.cm,angle=270.]{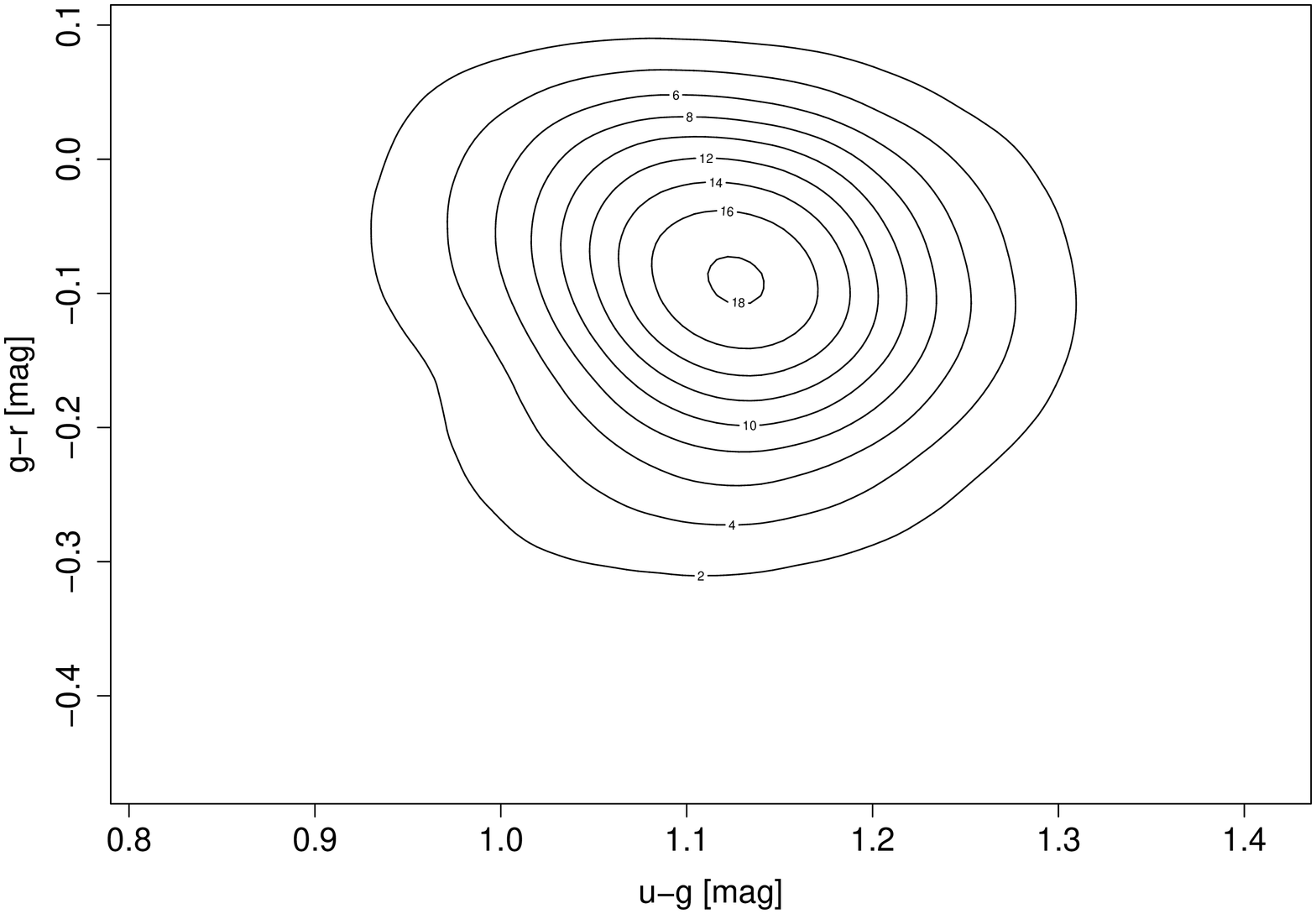}
\includegraphics[height=9.cm,angle=270.]{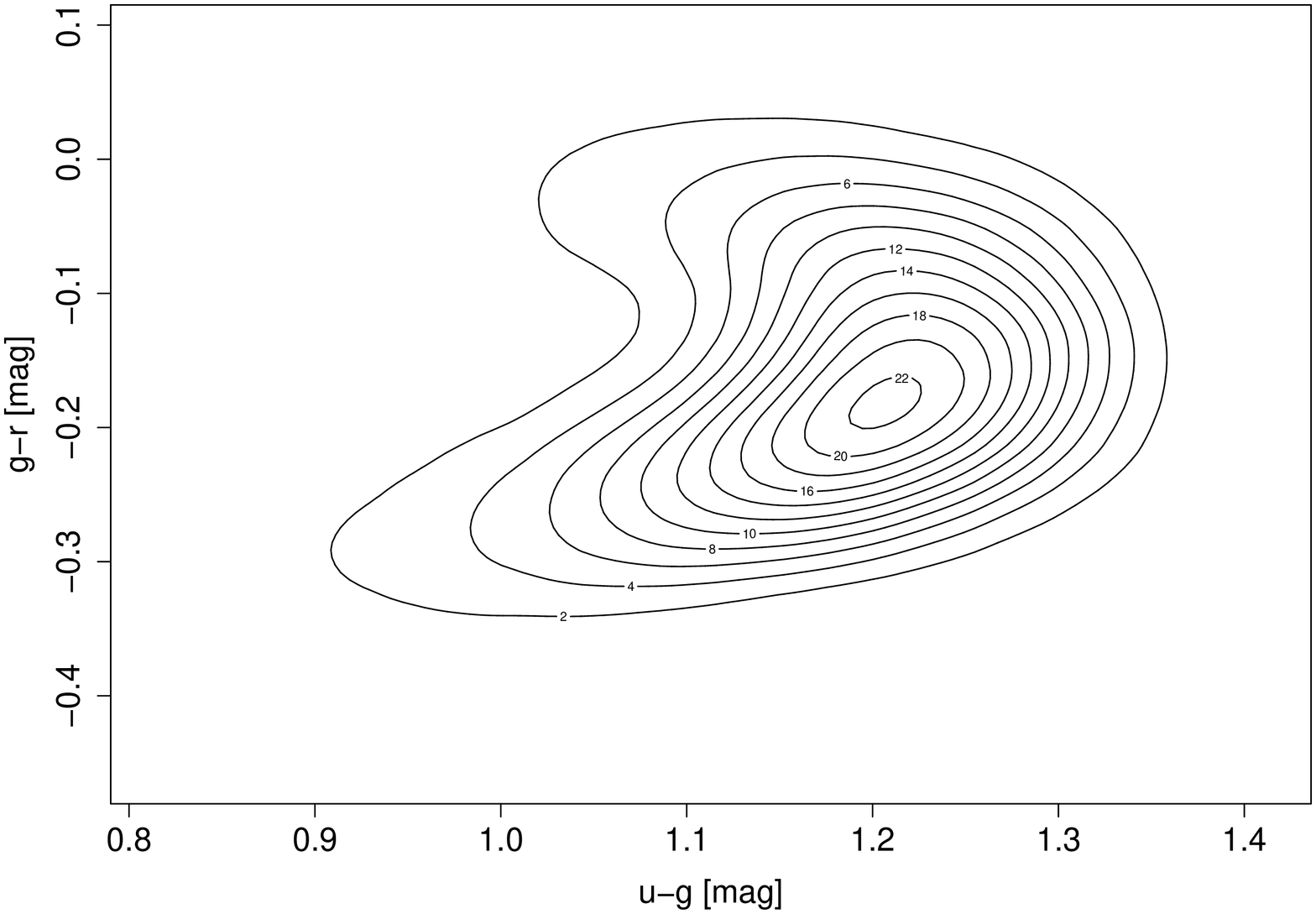}
}
\end{center}
\caption{The density of points for the non-BHB star training set (left)
and the BHB star training set (right) in the $u-g$,~$g-r$ plane. 
These density functions are used for the KDE classification. Contours range 
from 2 to 22 stars per unit area in steps of 2. 
\label{kde_ugr_dens}}
\end{figure*}

Trial and error experimentation with the available colours shows that
reasonable results can be obtained with $u-g$ and $g-r$, but the addition of
further colours degrades the performance.  
We construct density functions for both
the BHB stars and the non-BHB stars and compare the values at the locations of
test data or new data points to classify the source. The individual density
functions for BHB and non-BHB sources in the training set are shown as
contours in $u-g$, $g-r$ space in Fig.~\ref{kde_ugr_dens}.  The probability
of an object being of class $c1$ from a number $n_c$ of possible classes is
taken to be
\begin{equation}
P(j)=\frac{K_{j=c1}({\bf x})} { \sum_j^{n_c}{K_j({\bf x})}}
\label{kdeprobeqn}
\end{equation}
where $K_j$ are the density functions for each class
and ${\bf x}$ are the data.  

Training and test sets were independently selected ten times and used to train
and test a model.  We applied the KDE classifier to the test set under the
assumptions of equal class sizes (flat prior), the simple prior ($P(BHB)=0.32$
for all sources), and the 2d prior.   The results for all these
tests are shown in Table~\ref{kde_perf_1}. The layout of this table is the
same as Table~\ref{knn_basic}.
\begin{center}
\begin{table}
\caption{Results for KDE classification.
\label{kde_perf_1} }
\begin{tabular}{lll} 
Equal priors  & & \\
\multicolumn{3}{c}{Absolute} \\
        &  bhb     &  other  \\
BHB     &  1021  &  222   \\
OTHER   &   505  &  713   \\ 
\multicolumn{3}{c}{Percent} \\
        &  bhb     &  other  \\
BHB     &   82.1 \% &  17.9 \%   \\
OTHER   &   41.5 \% &  58.5 \%   \\
        &           &            \\
        &   Completeness &  Contamination \\
BHB     &    0.821            &  0.512     \\
        &                  &              \\    \hline
With simple prior & & \\   
        &     bhb        &   other     \\
BHB     &     765        &    478      \\
OTHER   &     281        &     937      \\
\multicolumn{3}{c}{Percent} \\
BHB     &     61.5 \%      &    38.5 \%   \\
OTHER   &     23.1 \%     &    76.9 \%   \\ 
        &                 &                \\
        &  Completeness   &  Contamination \\
BHB     &    0.615        &     0.438   \\
        &                 &             \\ \hline
With 2d prior &           &             \\
        &     bhb        &   other     \\
BHB     &     912        &    331      \\
OTHER   &     293        &    925      \\
\multicolumn{3}{c}{Percent} \\
BHB     &     73.4 \%      &    26.6 \%   \\
OTHER   &     24.1 \%       &   75.9 \%   \\ 
        &                 &                \\
        &  Completeness   &  Contamination \\
BHB     &    0.734        &     0.405   \\
\end{tabular}
\tablefoot{
{\em Top section:} Confusion matrix showing 
results of KDE classification
of 10 independently selected test sets based on 10 independently trained 
models. The classification threshold is $P(BHB) \ge 0.5$ in each case. 
The results are shown at the top as mean numbers of 
sources in each category and then as percentages. The 
completeness and contamination for the BHB output sample is also shown. 
These are corrected for the 
expected unbalanced class fractions.
{\em Middle section:} 
The same quantities calculated with the simple prior applied to the 
classifier output probabilities.
{\em Bottom section:} 
The confusion matrix, BHB completeness and BHB contamination obtained when the 
prior as a function of $g$ and latitude is applied to the classifier 
output probabilities. }
\end{table}
\end{center}

\begin{figure}[htbp!]
\begin{center}
\includegraphics[height=8.cm,angle=270.]{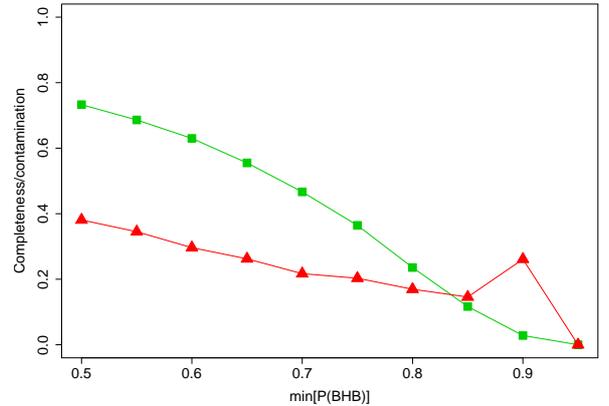}
\end{center}
\caption{Effect of varying the probability threshold with the KDE method for
  classification as a BHB star on the completeness (green squares) and
  contamination (red triangles) of the output. The results shown in
Table~\ref{kde_perf_1} correspond to a probability threshold of 0.5.
Output probabilities have been modified by the 2d prior. 
\label{kde_threshold_plot}}
\end{figure}

As with the kNN method, we experimented with thresholds at different levels of
classification confidence. We adjust the threshold probability for BHB
classification and record the resulting output
sample completeness and the contamination.  These are shown in
Fig.~\ref{kde_threshold_plot}. As expected, the effect of introducing a
threshold higher than 0.5 for classification is to reduce both the
completeness and contamination. 

\begin{figure}[htbp!]
\begin{center}
\vbox{
\includegraphics[height=8.cm,angle=270.]{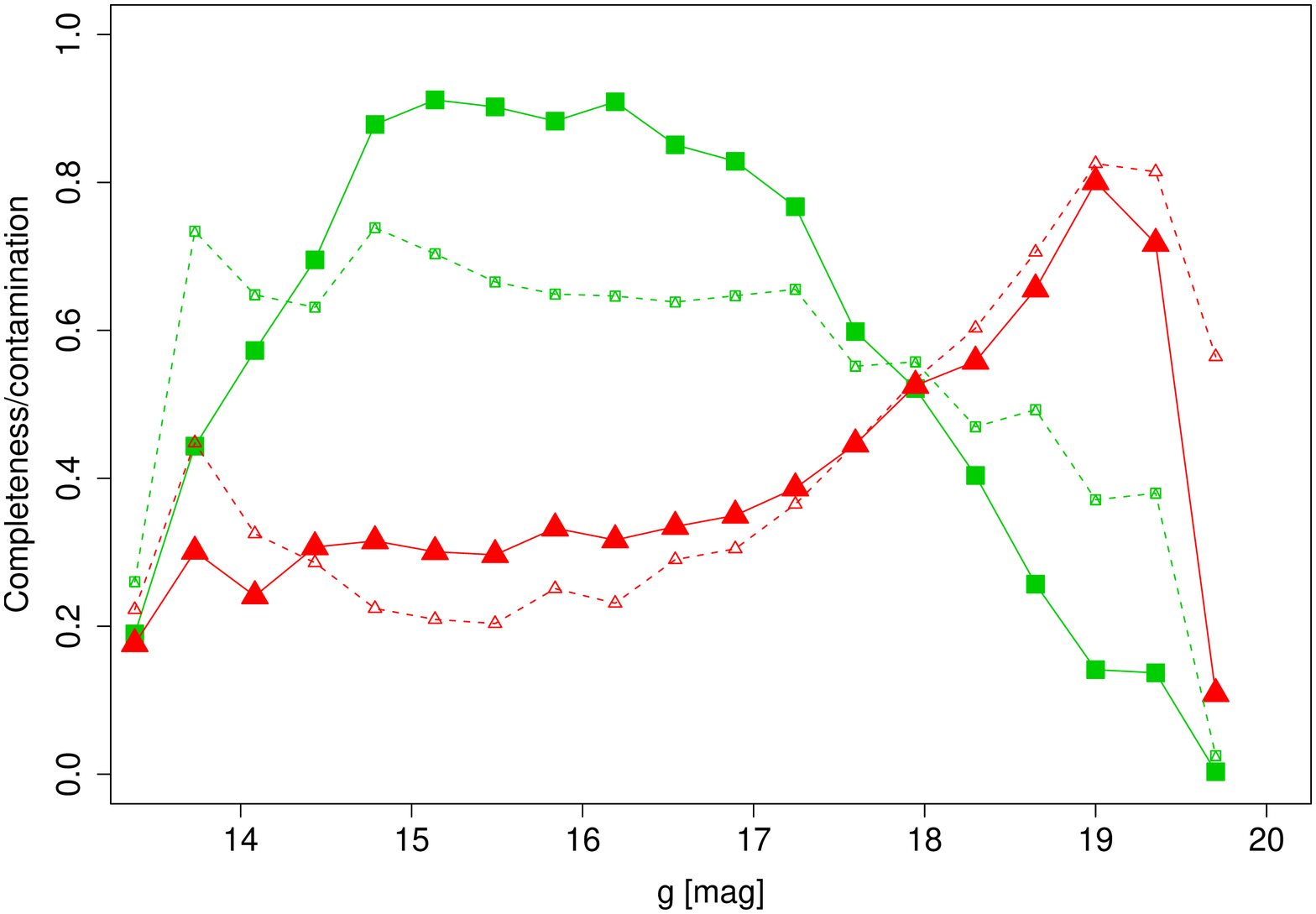}
\includegraphics[height=8.cm,angle=270.]{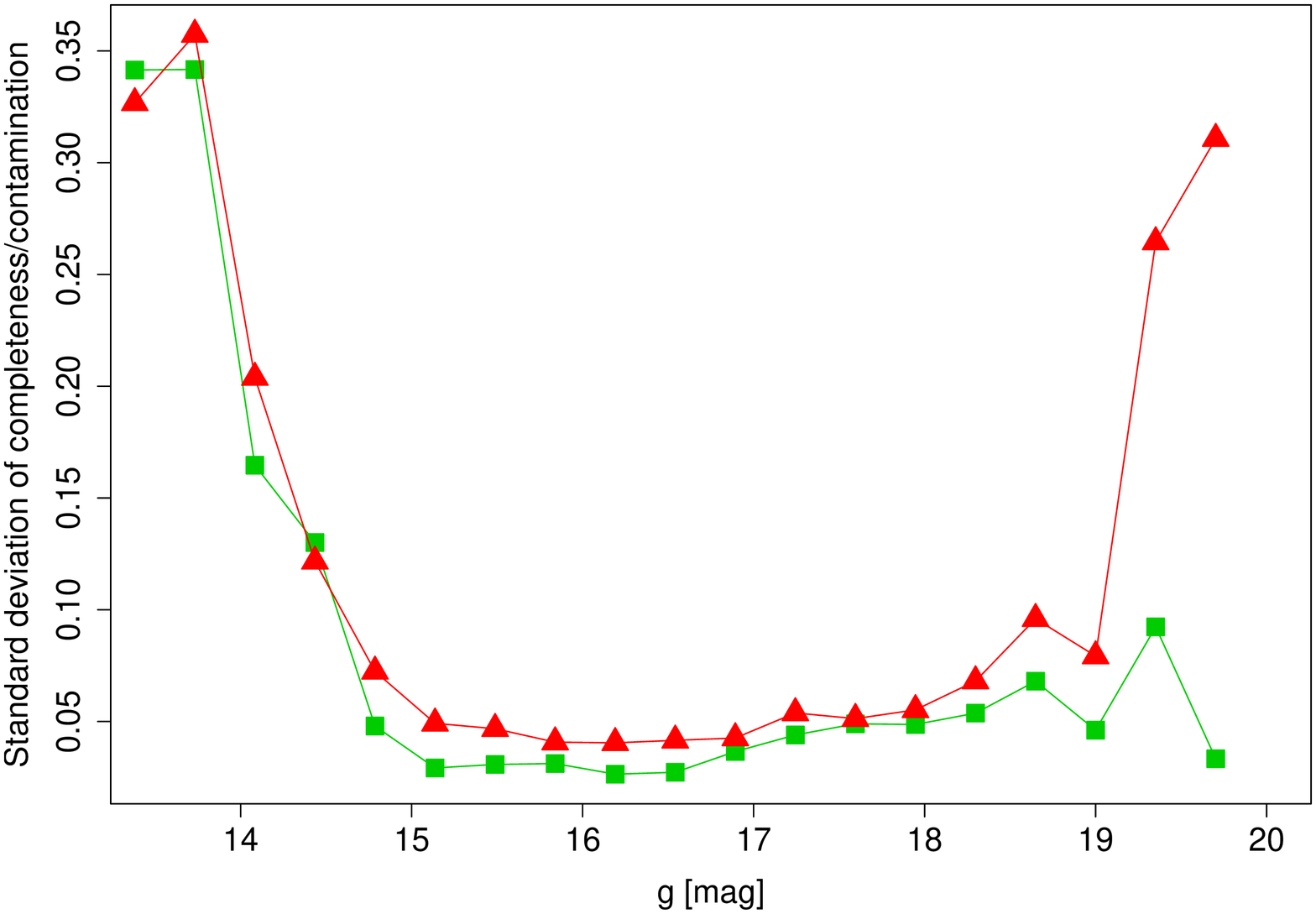}
}
\end{center}
\caption{{\em Top:} Completeness (green squares) and contamination (red triangles) 
for sources of different magnitudes classified with the KDE technique
with one hundred trials.  
The filled symbols and solid lines show the results using the 2d prior.
The open symbols and dashed lines show the results using the simple 
prior only.  
{\em Bottom: } Standard deviation of completeness and contamination over one hundred trials. 
\label{magdepcompcontamKde}}
\end{figure}

Figure~\ref{magdepcompcontamKde} shows the completeness and contamination 
for the test sample classified with the KDE method, with the results 
binned by magnitude. The threshold for classification is always 0.5. 

\subsection{Support vector classification}
\label{svms}

Support vector classification is a supervised method in which a high
dimensional decision boundary is fit between two classes. 
The boundary is chosen to maximize
the margins with the nearest representative points of each class (the
so-called support vectors). 
See \cite{vapnik} for a fuller description.
A linear SVM defines a boundary that is linear in
the original data space (in our case the 
four SDSS colours). By using a kernel
function, a higher dimensional feature space can be defined, and the
decision boundary instead defined in this. We use the second order radial
basis function as a kernel here. This function has a single parameter, gamma,
which must be set before training the model. To deal with the problem of
regularization for noisy data, a cost parameter can be introduced, that acts
to soften the margin. The cost parameter is so called because it controls the
extent to which the algorithm will attempt to fit a more complex boundary in
order to correctly classify all of the training points, i.e. it is the 'cost'
to the algorithm of misfitting training points during the model training
\citep{cortesvapnik}. We use the libSvm implementation,
which is available online (\url{http://www.csie.ntu.edu.tw/~cjlin/libsvm/}, \citet{changlin}) 
and is implemented in the R package e1071.

\subsubsection{Probabilities from SVM}

The SVM method is not designed to provide probabilities, since it deliberately
discards many of the training points, using only the support vectors to build
the model of the decision boundary. However, a probability estimate can be
made based on the distance of a test point from the decision boundary
\citep{platt}. The actual probability returned is based on a model fitted to
the training data. This probability estimate is essential if we want to trade
off completeness versus contamination, or use priors.

The training data were standardized colour by colour
so that each of the colours had zero mean and unit standard deviation.  The
same offset and scaling, calculated from the training data, 
were applied to the testing data. The SVM was run over
a grid of parameters; cost and gamma, with a fourfold cross validation using
the training data to determine the best choice for these values.  The optimum
values chosen were gamma=0.25, cost=64. The model was then trained on the
training set and applied to the test set. The basic classification performance
is shown in Table~\ref{svm_perf_1}.

\begin{figure}[htbp!]
\begin{center}
\includegraphics[height=8.cm,angle=270.]{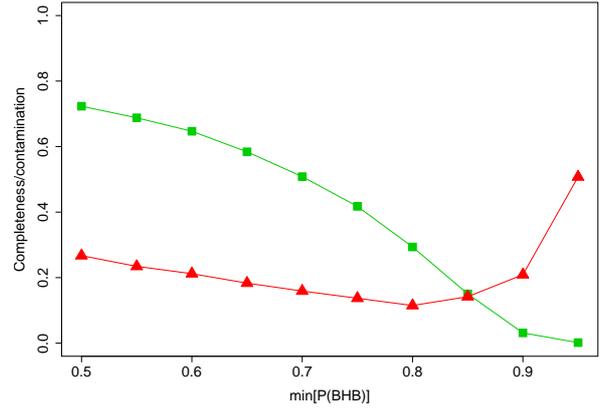}
\end{center}
\caption{
Plot of completeness (green squares) and contamination (red triangles)
as a function of a threshold probability for BHB classification 
in the case of the SVM classifier. 
The results shown in
Table~\ref{svm_perf_1} correspond to a probability threshold of 0.5.
Output probabilities have been modified by the 2d prior. 
\label{svm_thresholds}}
\end{figure}

\begin{center}
\begin{table}
\caption{Results for SVM classification.
\label{svm_perf_1} }
\begin{tabular}{lll} 
Flat prior   &      & \\
\multicolumn{3}{c}{Absolute} \\
        &  bhb   &  other  \\
BHB     &   988  &   255   \\
OTHER   &   305  &   913   \\ 
\multicolumn{3}{c}{Percent} \\
        &  bhb     &  other  \\
BHB     &   79.5\% &  20.5\%   \\
OTHER   &   25.0\% &  75.0\%   \\
        &           &          \\
        & Completeness & Contamination \\
BHB     &   0.795      &   0.396 \\
        &              &          \\ \hline
Simple prior  & & \\
\multicolumn{3}{c}{Absolute} \\
        &  bhb     &  other  \\
BHB     &  836     &  407  \\
OTHER   &  188     &  1030  \\
\multicolumn{3}{c}{Percent} \\
BHB     &  67.3\% &  32.7\% \\
OTHER   &  15.4\% &  84.6\% \\
        &         &         \\
        &  Completeness & Contamination \\
    BHB & 0.673    & 0.323 \\
        &         &      \\ \hline
With 2d prior & & \\
\multicolumn{3}{c}{Absolute} \\
        &  bhb     &  other  \\
BHB     &  912     &  331  \\
OTHER   &  187     &  1031  \\
\multicolumn{3}{c}{Percent} \\
BHB     &  73.4\% &  26.6\% \\
OTHER   &  15.4\% &  84.6\% \\
        &         &         \\
        &  Completeness & Contamination \\
    BHB & 0.73    & 0.303 \\
\end{tabular}
\tablefoot{
{\em Top section:} Confusion matrix showing results of an 
SVM classification 
without priors and with a threshold of $P(BHB)>0.5$, 
also the completeness and contamination in the output 
BHB sample, corrected for the expected class imbalance. 
{\em Middle section:} the same quantities obtained 
applying the simple prior 
probability for all sources. 
{\em Bottom section:} The confusion matrix and completeness and contamination 
found when applying the prior as a function of $g$ and latitude. }
\end{table}
\end{center}
We consider the effect of a threshold on the measured completeness and
contamination. The results of introducing various thresholds greater than 
$P(BHB)=0.5$ are shown in Fig.~\ref{svm_thresholds}.
It can be seen from Fig.~\ref{svm_thresholds} that the completeness and
contamination both fall as the threshold is increased, except for very high
thresholds when the contamination in fact rises. This is possible if the set
of sources with the highest values of P(BHB) contain a large number of
contaminants. This is undesirable, but is partly caused 
by the low number of sources with high 
$P(BHB)$ - in fact there are only thirteen sources with $P(BHB)>0.9$.

\begin{figure}[htbp!]
\begin{center}
\vbox{
\includegraphics[height=8.cm,angle=270.]{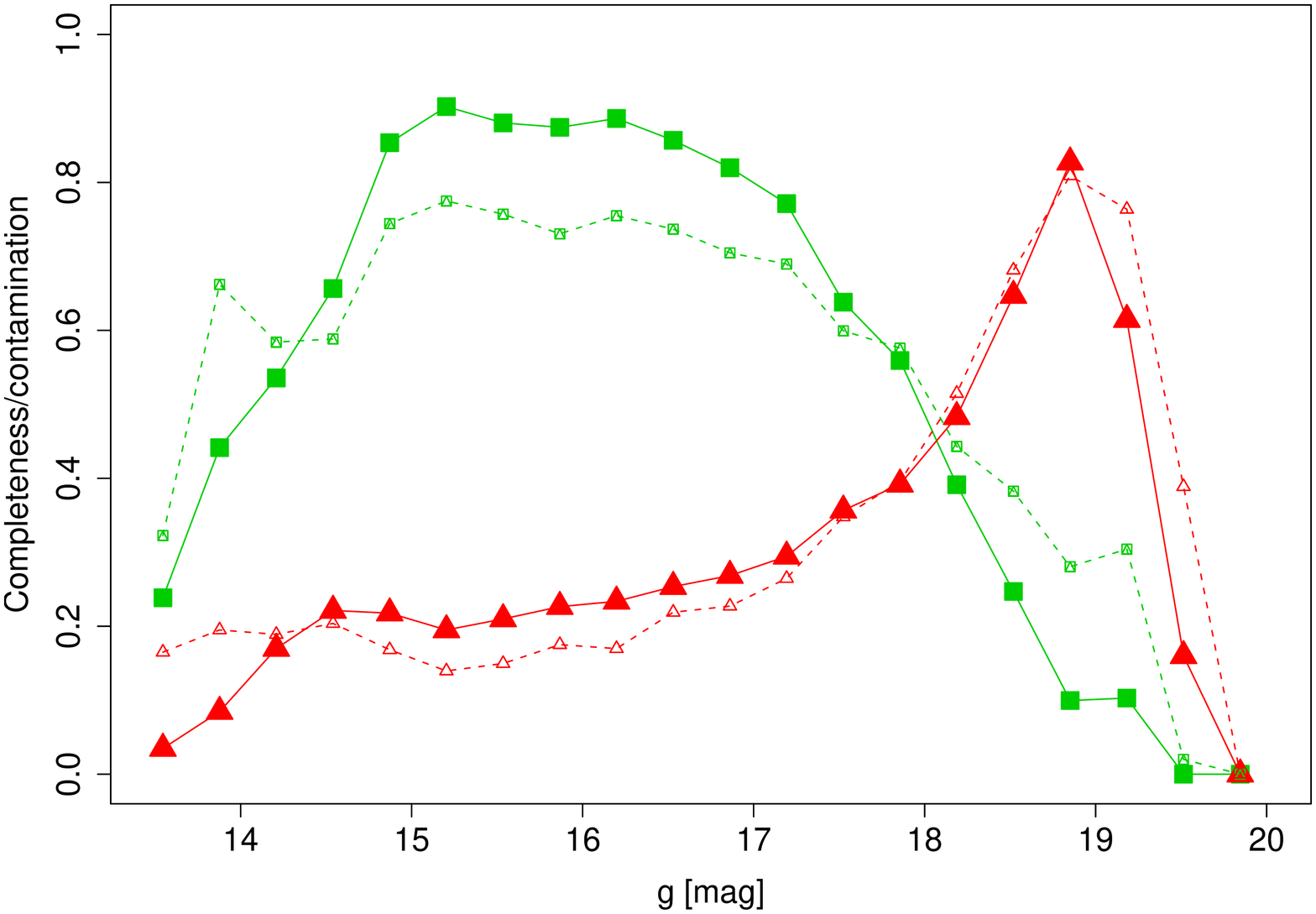}
\includegraphics[height=8.cm,angle=270.]{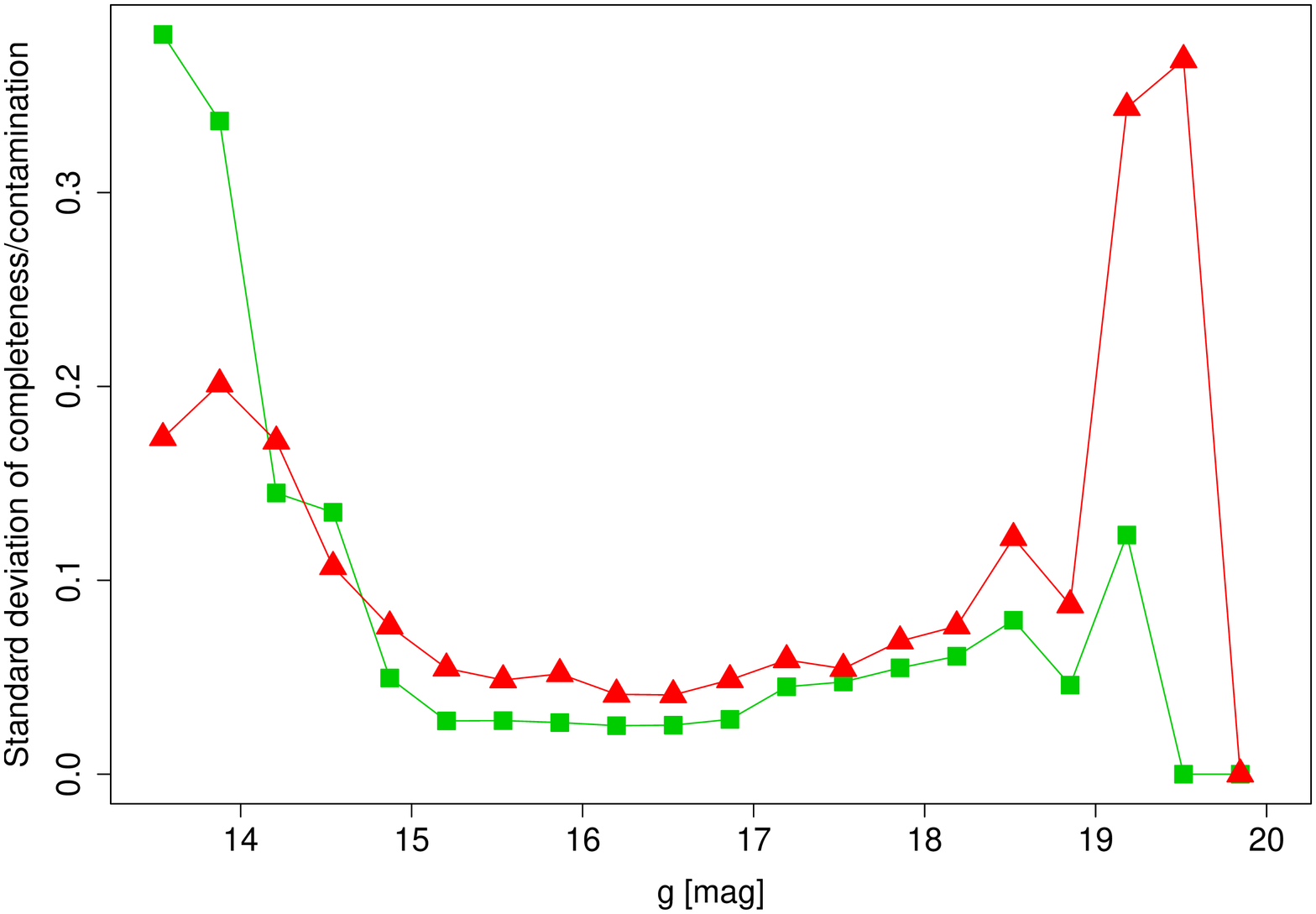}
}
\end{center}
\caption{{\em Top:} 
Completeness (green squares) and contamination (red triangles) 
for sources of different magnitudes classified using SVM with one hundred 
trials.  
The filled symbols and solid lines show the results using the 2d prior.
The open symbols and dashed lines show the results using the 
simple prior only.  
{\em Bottom: }Standard deviations of one hundred trials.
\label{magdepcompcontamSvm}}
\end{figure}

Figure~\ref{magdepcompcontamSvm} shows the completeness and contamination 
for the test sample classified with SVM, with the results 
binned by magnitude. This plot shows the results both with the 
simple prior and the 2d prior. As with the KDE method, the SVM performs well 
for $14<g<18$ but progressively more poorly for fainter sources.   

\subsection{Optimal choice of classifier}

From the bare results in Tables~\ref{knn_basic},~\ref{kde_perf_1}
and~\ref{svm_perf_1}, using the 2d prior, all the techniques have very similar
completeness. The contamination is best in the  case of SVM with about 0.3, and
worst for the KDE with 0.4. 
\begin{figure}[htbp!]
\begin{center}
\vbox{
\includegraphics[height=8.cm,angle=270.]{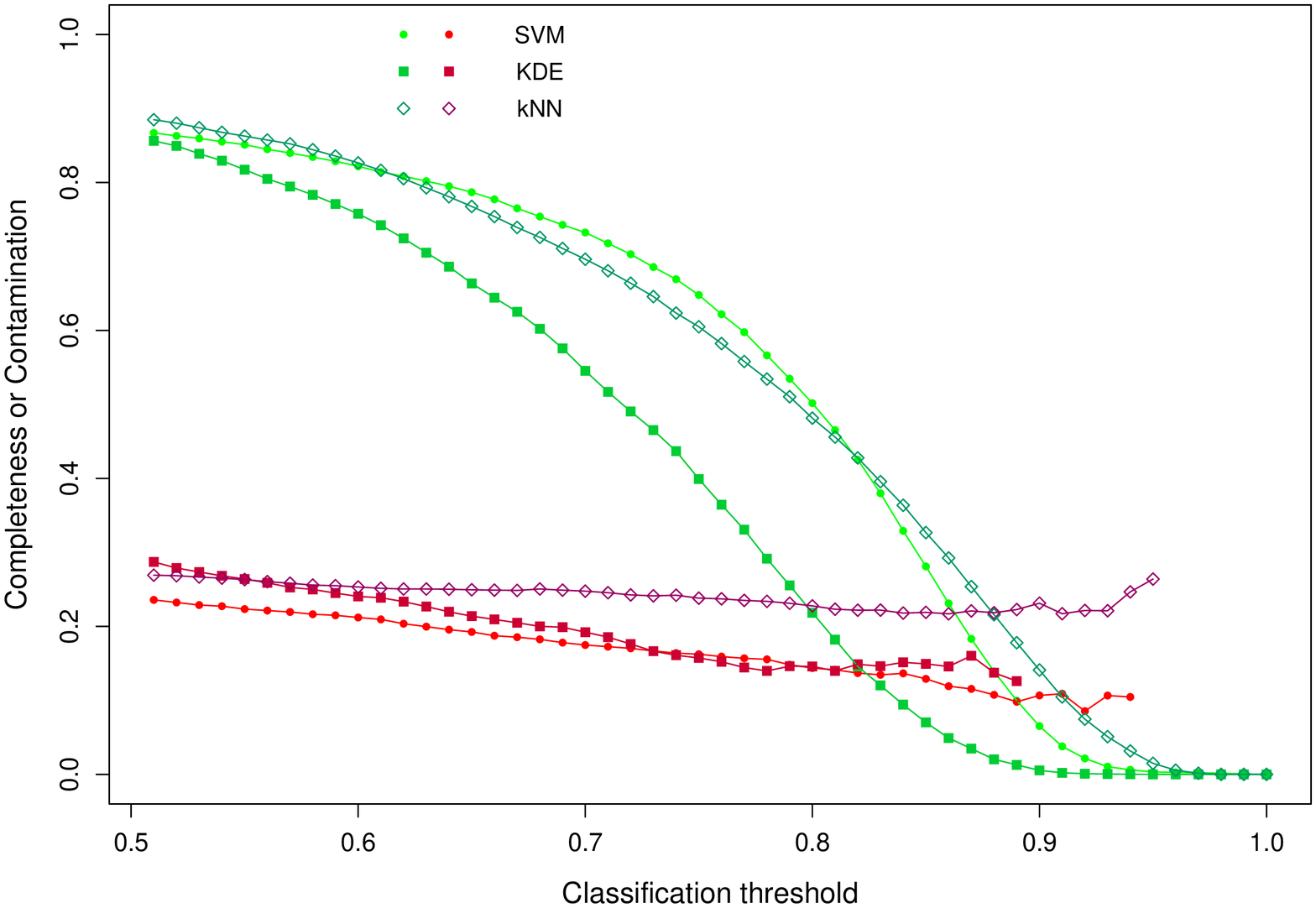}
\includegraphics[height=8.cm,angle=270.]{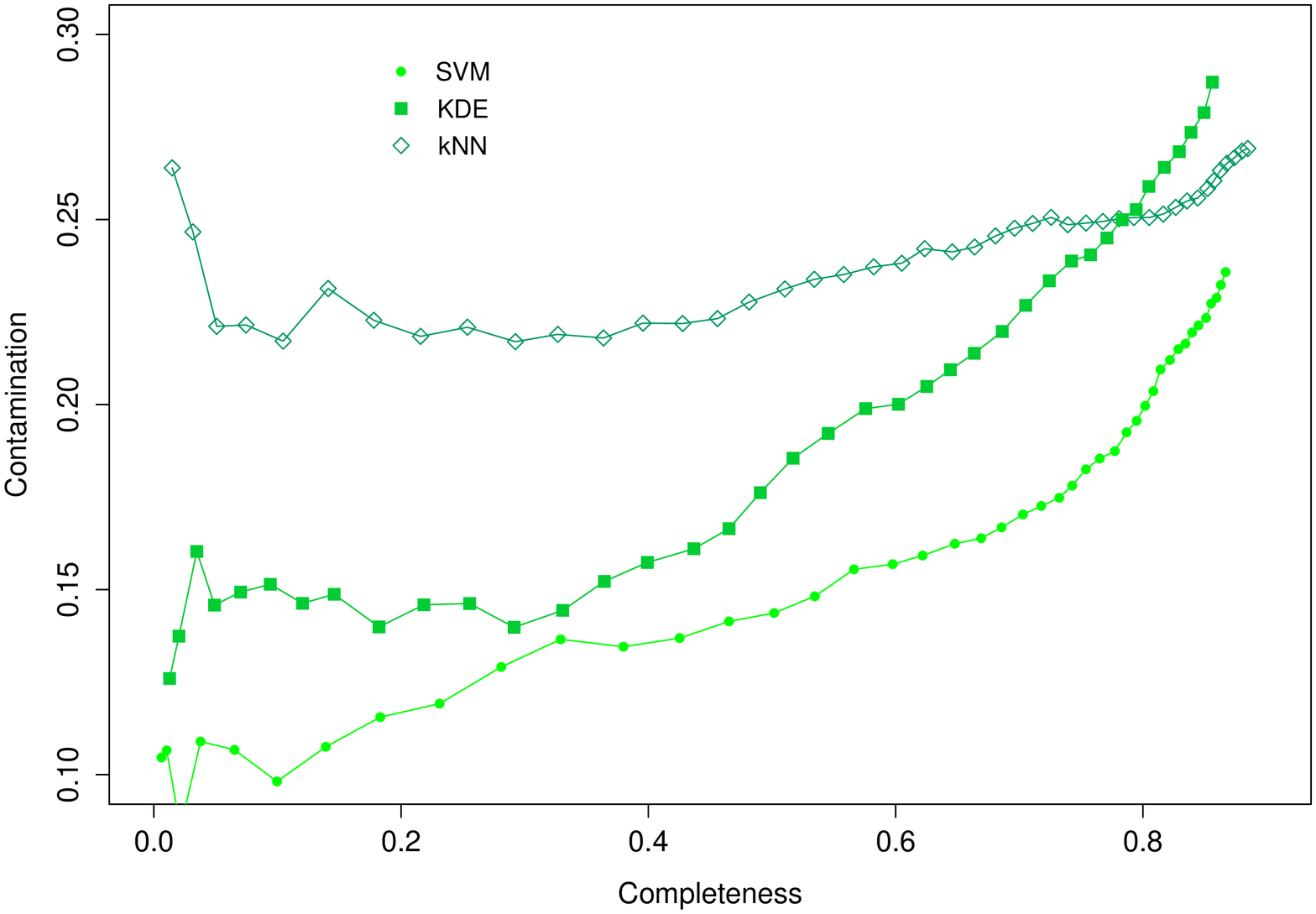}
}
\end{center}
\caption{{\em Top:}
Completeness (green) and contamination (red) 
for test samples with $15<g<17$ classified with the SVM, KDE or kNN. 
Different shades and symbols are used to distinguish the
methods. All results are modified with the 2d prior and 
the contamination is 
corrected for class fractions. 
Results are the average of 
ten independent runs. {\em Bottom:} Same data as in the top plot, with 
completeness plotted directly against 
contamination for direct comparison. The SVM results are always below 
and to the right of the other methods, demonstrating lower contamination for 
a given completeness.
\label{compareResults}}
\end{figure}
 The decision boundary method of \cite{yanny} should properly be compared
  with the simple prior case for the three machine learning methods -- the
  test set naturally has the right class fractions, but because the method is
  not probabilistic, the correction for the 2d prior cannot be made.  The
  completeness of the decision boundary method is clearly better than the
  three machine learning methods. The contamination of over 50\% is however
  worse than any of them.

From the magnitude performance plots in
Figs.~\ref{magdepcompcontamKnn},~\ref{magdepcompcontamKde}
and~\ref{magdepcompcontamSvm}, it can be seen that all the methods achieve a
high completeness and low contamination for the approximate range $15<g<17$.
The contamination
achieved by the SVM technique for the region of best performance between 
$15<g<17$ is
slightly better than for the kNN.

To make a more direct comparison, we plot in Fig.~\ref{compareResults}
the completeness and contamination, averaged over ten independent trials, 
for all the methods as a function 
of the classification threshold. For this plot, we restrict the 
test sample to sources in the range $15<g<17$, where all the methods 
perform reasonably well.

 From this comparison, we can note the following; The SVM and
  kNN methods deliver similar completeness over most of the range of
  thresholds. The kNN technique maintains completeness better than SVM for
  very high thresholds. However, the kNN method does not show any significant
  improvement in contamination, and it
  never delivers a better contamination than the other methods for
  similar completeness.  The other techniques do show a falling contamination
  with increasing threshold. In the lower plot, it is
  clear that the SVM delivers on average a lower contamination for a given
  completeness.

In summary, all the methods perform reasonably well, but the SVM seems to have
the edge across the largest range of conditions, and we choose to use this
technique on the new data.

\section{Classification of new data}

\begin{figure}[htbp!]
\begin{center}
\includegraphics[height=8.cm,angle=270.]{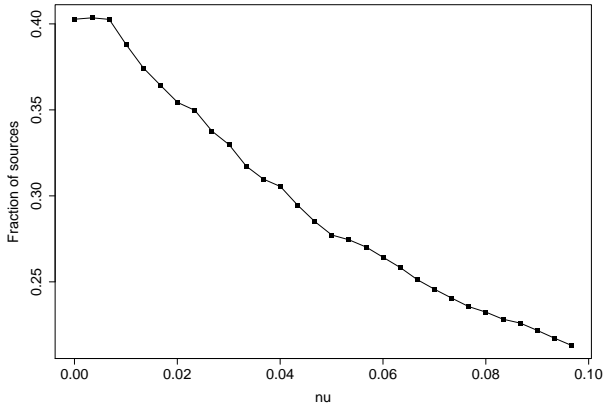}
\end{center}
\caption{The fraction of new data points passing through the one class filter as
a function of the parameter $\nu$. The higher the value of the parameter, the more objects are rejected, and fewer then remain for the main classification. 
The fraction reaches a plateau at around $\nu=0.01$.     
\label{oneClassFilterTest}}
\end{figure}

We obtained new DR7 photometry from SDSS and use the various models to predict
the classes (BHB versus non-BHB). The DR7 data were obtained using the colour
cut of \citeauthor{yanny} (see Sect.~\ref{sect:data}).  The search yielded
859,341 objects at $g<23$.  This magnitude cutoff is very deep, being 0.8
magnitudes deeper than the 95\% completeness limit for DR7
\citep{dr7paper}. However, the selection requires good photometry in all five
SDSS bands, and the classification method will enforce the condition that
classified objects occupy the data space defined by the training set (see next
section), so that the number of spurious objects in the sample will eventually
be very low.
\begin{figure}[hbtp!]
\begin{center}
\vbox{
\includegraphics[height=8.cm,angle=270.]{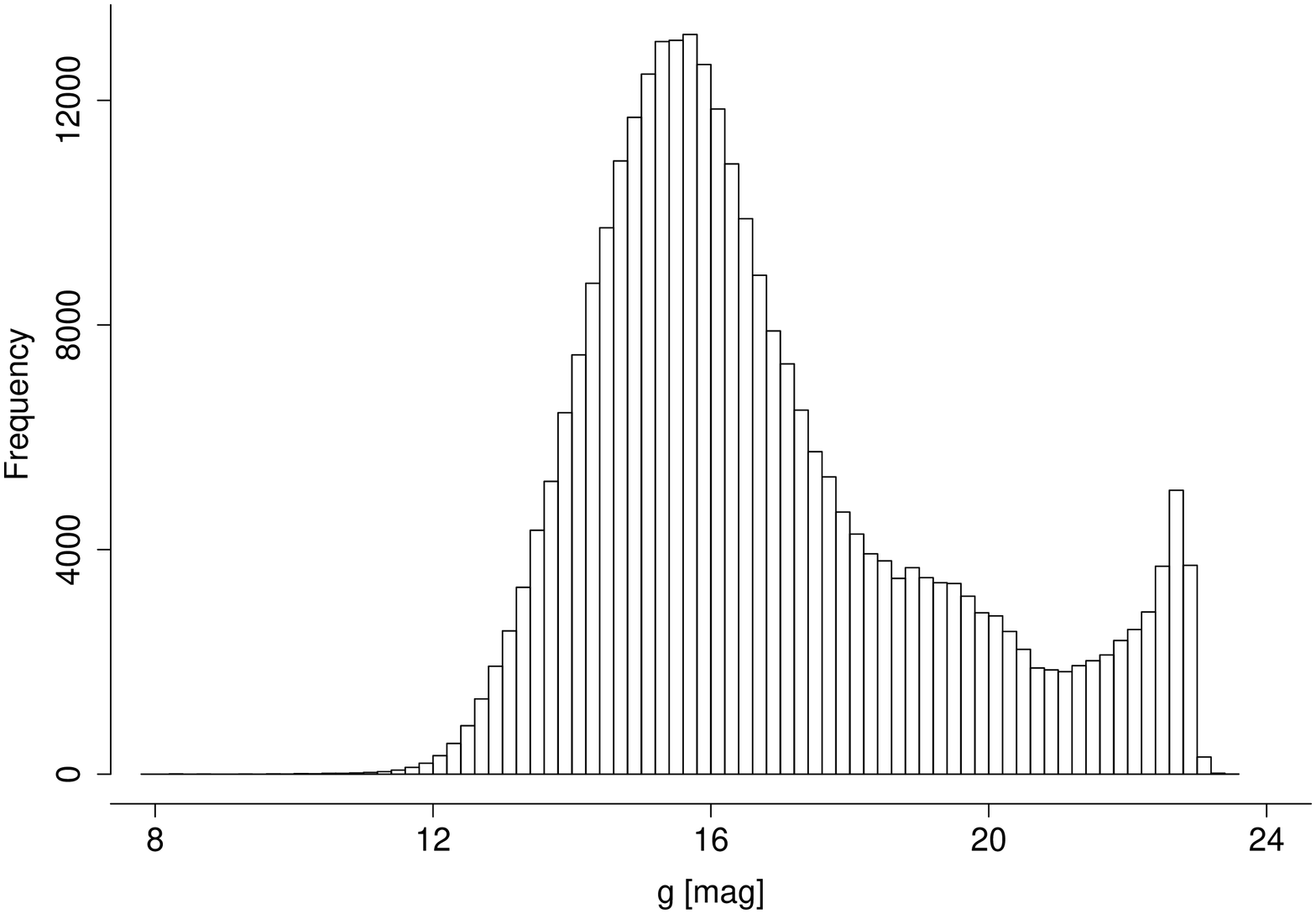}
\includegraphics[height=8.cm,angle=270.]{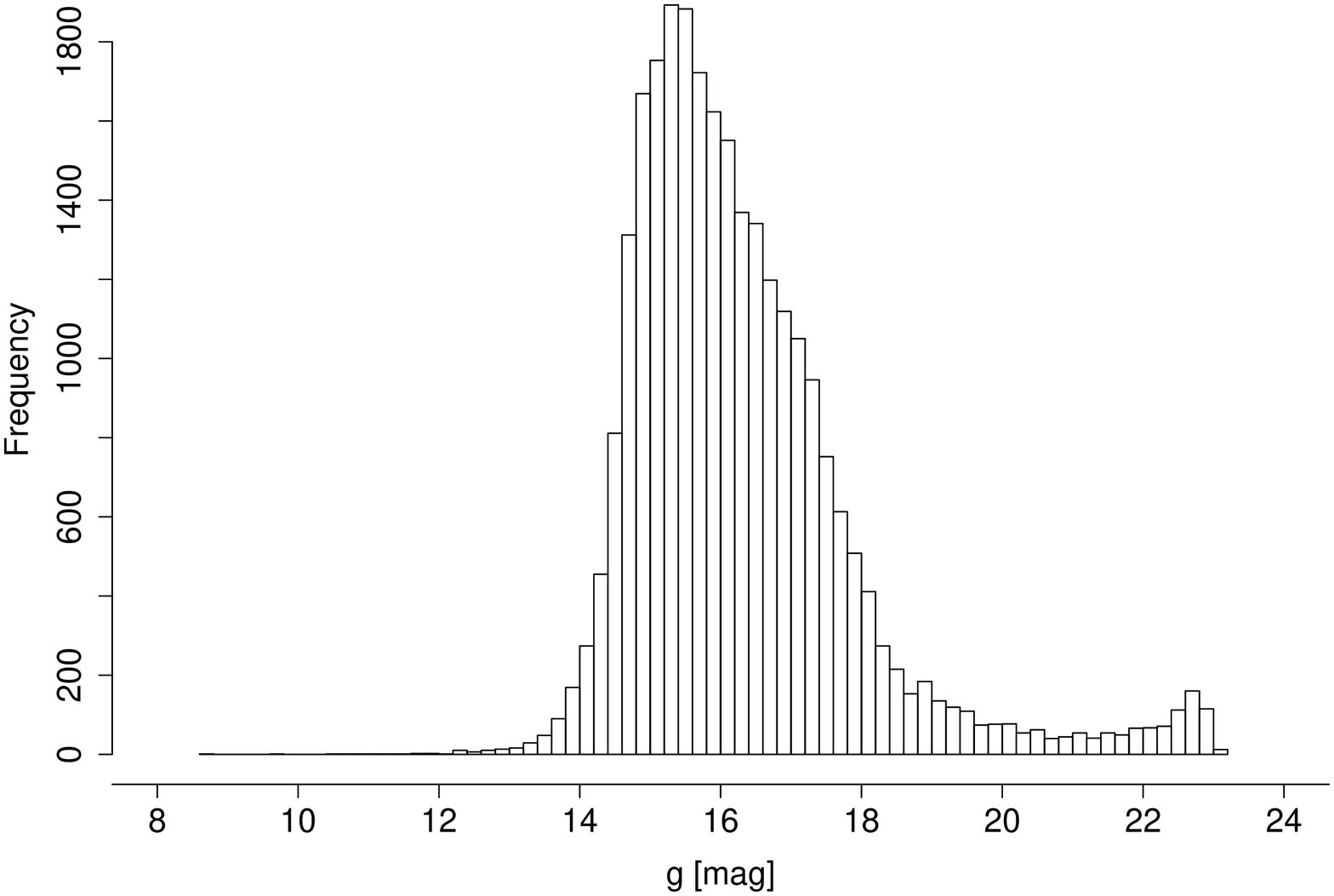}
}
\end{center}
\caption{({\em Top}) $g$ magnitude distribution of 
sources from DR7 selected to lie in the same 
data space as the SVM training data. ({\em Bottom}) $g$ magnitude distribution
of sources classified as BHB stars.  
\label{magnitudeDistributions}
}
\end{figure}

\subsection{One class filter}

 Before attempting to classify the new data, it is necessary to exclude
  points which lie outside the locus of the available training points.  This
  issue did not arise when using the testing data as described previously,
  since all input sources are by definition part of the defined data set and
  could potentially be used to train a classifier. New points lying in a
  'hinterland' outside the training data locus and well away from the decision
  boundary may be misclassified with high confidence levels, since the
  probability model is based on distance from the decision boundary. It is
  necessary to exclude such points prior to attempting the classification.

To do this, we used an SVM in one-class mode.
The one-class SVM defines a decision boundary which separates the training
data from the origin with a maximized margin.  A parameter, $\nu$, controls
the rigidity of the boundary, and hence what fraction of the training set
would typically be excluded. We collected all the available training objects
(2,536 BHBs and 7,511 non-BHBs) together into one set and standardized according
to this data. We conducted an experiment with different values of the
parameter $\nu$ to see how many of the new data points would pass through. The
results are shown in Fig.~\ref{oneClassFilterTest}
From this figure, we see that the fraction of sources passing through the 
filter reaches a plateau for values of $\nu$ a little less than 
0.01. We chose $\nu=0.01$ based on this fact, and on a visual 
inspection of the region of the data 
space in $u-g$,~$g-r$ space occupied by the surviving 
points. 

Filtering the
photometry according to consistency with the training set, we were left with
294,652 objects. The distribution of these in $g$ magnitude is shown in
Fig.~\ref{magnitudeDistributions} (left hand side).
  
\subsection{Classification}

\begin{figure}[htbp!]
\begin{center}
 \includegraphics[height=8.cm,angle=270.]{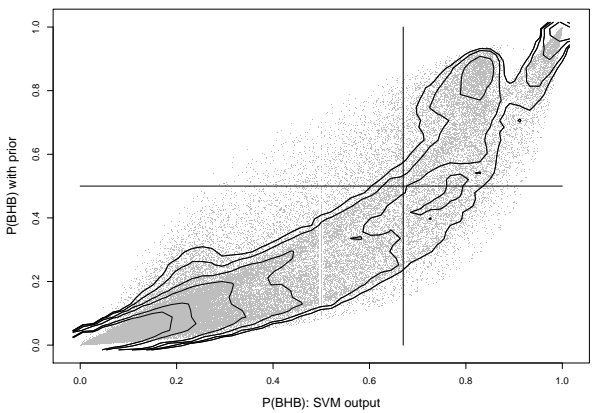}
\end{center}
 \caption{Classification probabilities P(BHB) for the new sources in the DR7
   data set. The abscissa shows the probability returned by the SVM
   classifier, the ordinate shows the probability after modification with the
   two dimensional prior (a function of latitude and $g$ magnitude).  The
   horizontal line marks the $P=0.5$ threshold above which sources were
   classified as BHB stars for the purposes of this work. The vertical
   line shows the equivalent threshold for the SVM raw probabilities assuming
   a simple prior of $P(BHB)=0.32$ for all sources. Contours are at 0.004,
   0.006, 0.01, 0.03, 0.09 and 0.12 times the maximum density. The
     highest density regions lie in the bottom left, top right and in the
     clump at approximately (0.8,0.8).
 \label{newDataProbs}
}
 \end{figure}

For the classification, we trained a new SVM model using all the available BHB
stars, plus an equal number of randomly selected non-BHB stars.  The 2d prior
probability was used to obtain posterior probabilities, and a threshold of 0.5
was applied to these. With this threshold, 27,074 of the new sample
objects were classified as BHB stars. Figure~\ref{newDataProbs} shows the
probabilities output from the SVM classifier plotted against the probabilities
modified by the 2d prior. The threshold for BHB classification is shown by the
horizontal line. The threshold obtained by assuming a prior probability
$P(BHB)=0.32$ for all sources is shown with the vertical line.  It can be
  seen that there are peaks in source density at low and high probability, so
that the choice of prior does not dominate the classification. The
distribution of classified BHB stars in $g$ magnitude is shown in the right
hand side of Fig.~\ref{magnitudeDistributions}.

\subsubsection{Stability of the probabilities}

The training set for the classifier is composed of all the available BHB
stars, together with an equal number of non-BHB stars selected at random.
This means in practice about one third of the non-BHB stars are included in
the training set. The output probabilities, and eventual classifications in
many cases, will eventually depend on the exact choice of training data. To
quantify the stability of the output probabilities, we performed ten
resamplings of the training data and subsequent classifications, and found the
standard deviations of the output probabilities. 

\begin{figure}[htp!]
\begin{center}
\includegraphics[height=8.cm,angle=270.]{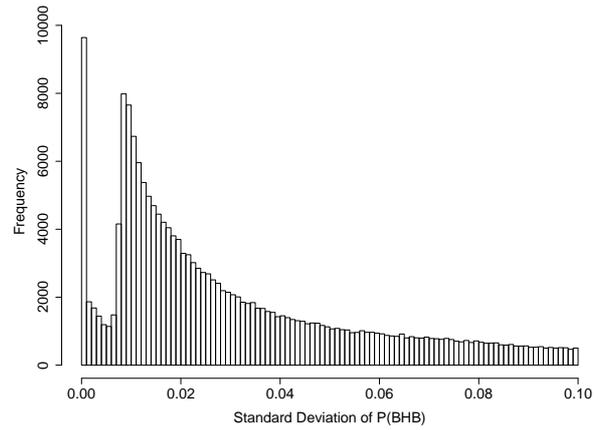}
\end{center}
\caption{Histogram of the standard deviations of the probabilities
  output by the SVM classifier over ten classifications trained on ten
  separate resamplings of the available training data. 
 \label{probabilityVariation}
}
 \end{figure}

In
Fig.~\ref{probabilityVariation} we show a histogram of the standard
deviations of the probabilities, and a plot of the standard deviation of the
probability for each output sources versus the mean value of the probability
obtained. There is a broad peak at about 0.015. The
histogram has been truncated at $P=0.1$.  

 In Table~\ref{probabilityVariationTable}, we show the percentages of objects
 classified as BHBs or non-BHBs with standard deviation in the probability
 exceeding 0.01, 0.05 and 0.1. The peak occurs between 0.01 and 0.02 (see also
 Fig.~\ref{probabilityVariation}). In each bin, fewer BHB sources than
 non-BHB sources have higher standard deviations than the given threshold.

\begin{table}
\caption{Probability standard deviations for BHB and non-BHB classes.
\label{probabilityVariationTable}}
\begin{center}
\begin{tabular}{lll}
SD $>$ & BHB   & non-BHB \\
0.01   & 72\%  & 77\%    \\  
0.02   & 34\%  & 55\%    \\
0.05   & 16\%  & 29\%    \\
0.10   & 10\%  & 14\%    \\              
\end{tabular}
\end{center}
\end{table}

It is difficult to combine repeated probabilities into a single value with an
uncertainty, and also to include the effect of the magnitude and latitude
dependent prior, and we do not attempt to do this. Instead, we give in the
output table (Table~\ref{catalogue}) the raw SVM output probability from one
classification, the standard deviation on this from ten resamplings, the
prior, and the posterior probability obtained by applying the prior to the raw
SVM probability. We use the condition that the posterior probability
$P(BHB)>0.5$ for BHB classification, but alternatively one could impose some
condition based on the standard error for each object.

\subsection{Photometric distances}
\label{sect:distances}

\cite{sirko} give absolute $g$ magnitudes based on models by
\cite{dorman93}, for a range of BHB star properties (Teff, logg and
metallicity - their Table 2), together with $u-g$, $g-r$ and $g-i$ colours .
To determine photometric distances for our BHB stars, we perform a regression
based on this data.  We do this with a support vector machine in regression
mode\footnote{The SVM for regression fits a regression line in a high
  dimensional feature space, rather than a classification boundary. This
  involves an extra parameter which must be tuned. For a full description see
  \cite{drucker:support} or the libSVM documentation.}.

\begin{figure*}[htbp!]
\begin{center}
\hbox{
\includegraphics[height=8.cm,angle=270.]{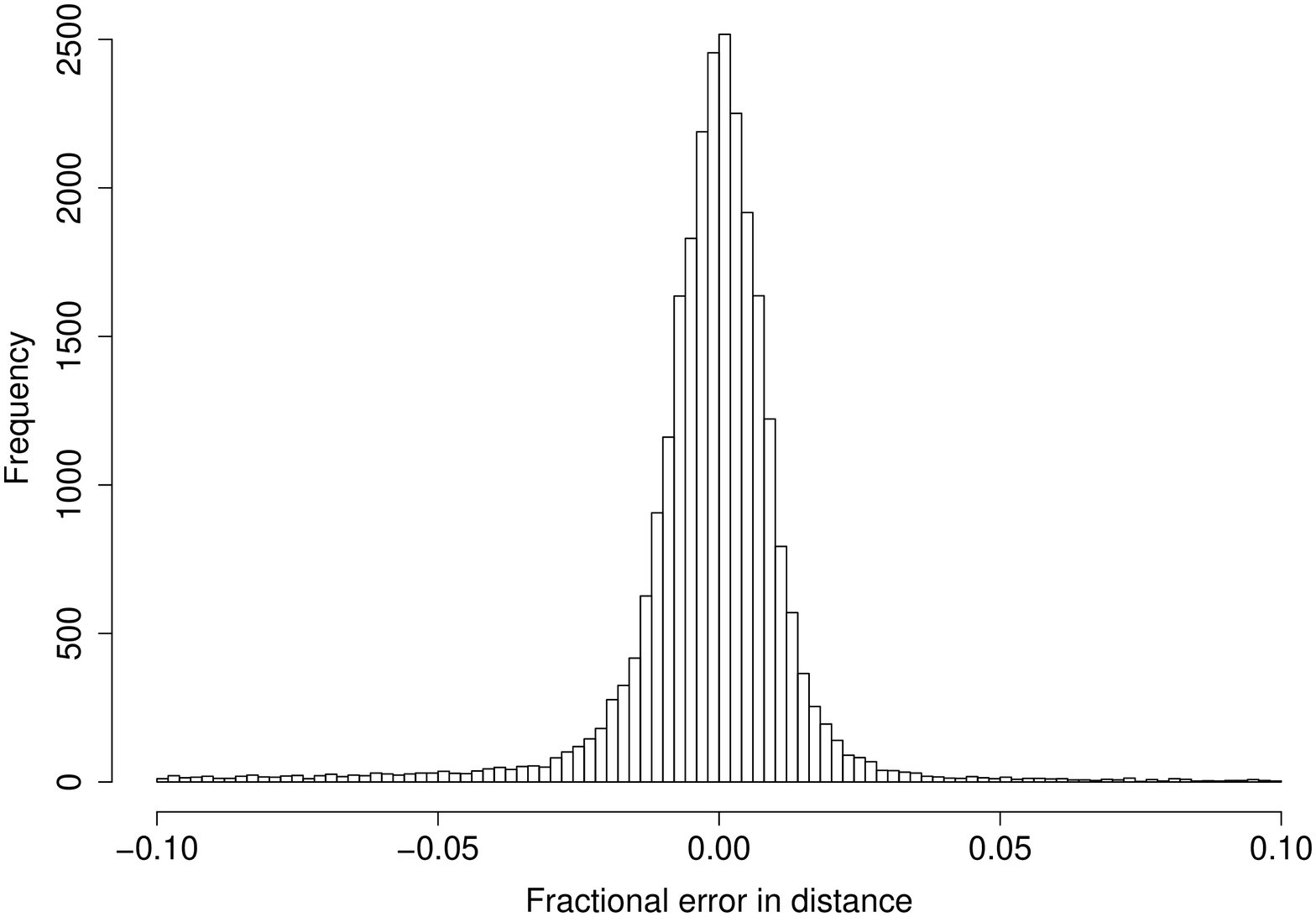}
\includegraphics[height=8.cm,angle=270.]{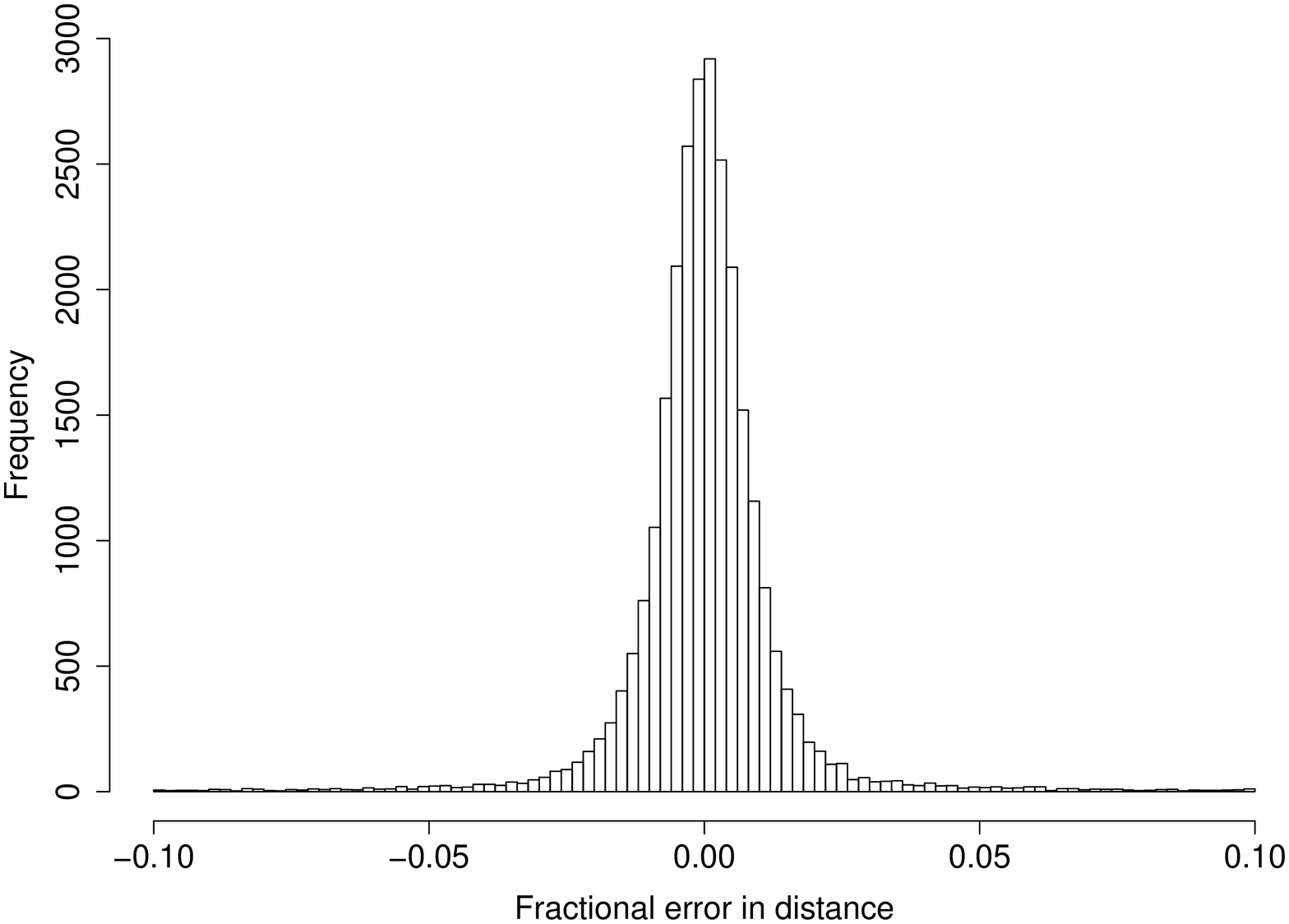}
}
\end{center}
\caption{The fractional change in the derived distance caused by 
applying a random offset to either the $g$ magnitude or the 
colours. The offsets are drawn from a Gaussian distribution 
with the appropriate $\sigma$ for each object. On the left, the effect of 
the colour uncertainty on the fitting result, on the right, the effect of the
photometric error in $g$. Both 
distributions have a width of the same order of magnitude.
 \label{distanceErrors}
}
 \end{figure*}

We estimate the distance errors due to the uncertainty in the photometry 
by recomputing the distances with $\pm 1\sigma$ in the colours and in 
$g$. the distributions are shown in Fig.~\ref{distanceErrors}.

\begin{figure*}[htbp!]
\begin{center}
\includegraphics[height=16.cm,angle=270.]{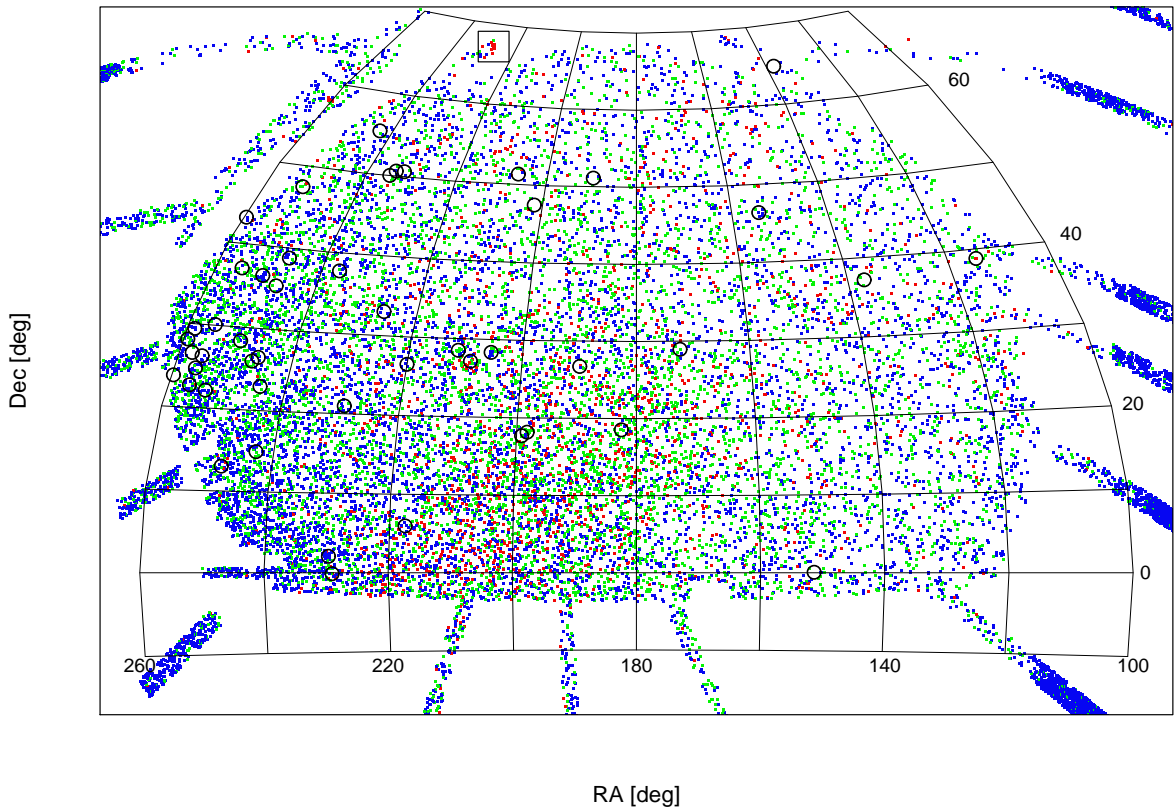}
\caption{BHB stars in the north Galactic cap region, shown in Aitoff 
projection. 
The stars have been colour coded for distance as follows; blue=closer 
than 15kpc, green=15-40kpc, red=further than 40kpc. 
The positions of a few globular 
clusters selected from the catalogue of \cite{harris} 
and used for a distance test 
are shown as black circles.  The position of the Ursae Minoris 
dwarf galaxy is marked with a box. 
\label{ngcPlusGcs}
}
\end{center}
\end{figure*}

Figure~\ref{ngcPlusGcs} shows the distribution of BHB stars on the sky in the
region of the north Galactic cap. The locations of a selection of globular
clusters taken from the list of \cite{harris} are shown as black
circles. These were used to make a test of the BHB distances, as described
below. Also indicated, with a box, is the location of the Ursa Minor
dwarf galaxy. The BHB population of this galaxy is clearly visible as a
clump of distant stars.

\subsection{A distance test}

\begin{figure*}[htbp!]
\begin{center}
 \includegraphics[height=14.cm,angle=270.]{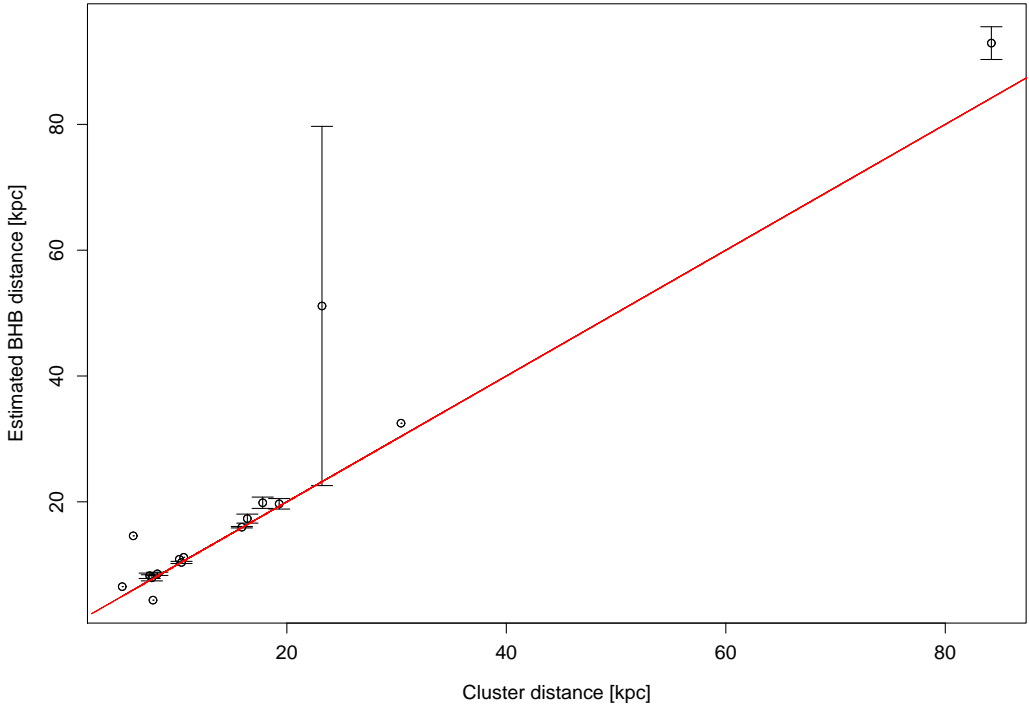}
 \caption{Distances to globular clusters taken from the catalogue of
   \citeauthor{harris} compared to the mean distance to BHB stars within one
   tidal radius of the cluster centre (both quantities taken from
   \citeauthor{harris}. The straight line shows exact agreement ($x=y$). Error
   bars are plotted where possible ($>1$ source identified). The main plot
shows the full sample, the inset shows the portion closer than 30kpc at a
larger scale. Globular clusters with no detected BHB population are not shown.  
\label{gcComparison}
}
\end{center}
 \end{figure*}

To help assess the accuracy of our BHB identifications and distance
determination, we compare our data to a selection of globular clusters taken
from the catalogue of \cite{harris} and selected to lie in the north Galactic
cap region. Their positions are shown in Fig.~\ref{ngcPlusGcs}. We
considered all BHB stars within half a tidal radius of each cluster centre
to be probable cluster members, 
and determined the mean distance to those stars, and the error on
the mean. We then compare those distances with the ones given in the table of
\citeauthor{harris}.  Out of fifty-two globular clusters within the north
Galactic cap region, sixteen had at least one BHB star from our sample within
half a tidal radius of the centre.  In Fig.~\ref{gcComparison} is shown the
mean distance for each cluster derived from the BHB population compared to the
accepted distance given in the catalogue. The agreements with the distances
taken from \citeauthor{harris} are generally reasonably good for clusters
around 20~kpc distant. Amongst the nearby clusters are several with
overestimated distances, two of which contain only one source
each. Overestimated distances would be expected if the cluster membership is
contaminated with non-BHB stars, since the contaminants are generally fainter
than the BHB stars and so the distances will be overestimated in those cases.
A more distant cluster at just over 80 kpc also has an overestimated distance.

\begin{figure}[h!]
\begin{center}
 \includegraphics[height=8.cm,angle=270.]{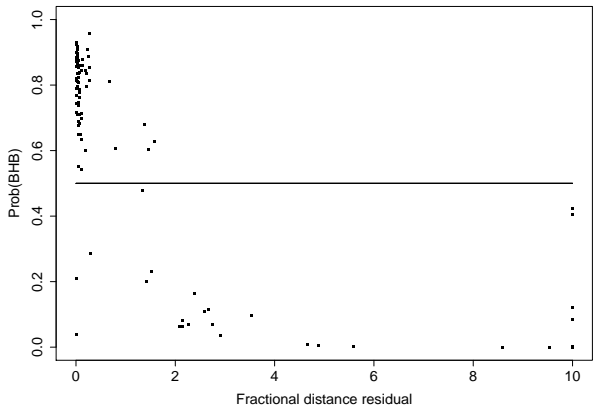}
\end{center}
 \caption{An analysis of classifier performance based on 
likely membership of known globular clusters. 
Sources are selected based on proximity to a known 
globular cluster from the catalogue of \cite{harris}. 
Distances are computed for all these sources, 
assuming they are all BHB stars, and compared 
to the catalogue distance for the cluster. 
True BHB stars that are really cluster members 
should return a distance consistent with the cluster 
distance. Most contaminants from within the cluster will 
appear too distant compared to the BHB population, 
because they are fainter. This plot shows the 
fractional absolute residual in the distance 
estimate in kpc versus the (prior-corrected) SVM probability $P(BHB)$.
The dashed line shows the threshold $P(BHB)=0.5$.            
 \label{distanceVsProb}
}
 \end{figure}

 In Fig.~\ref{distanceVsProb} we use the information from the
 globular clusters distance test to further investigate the performance of the
 classification. We select all sources within half a tidal
 radius of a cluster centre and calculate the distance to these assuming they
 are BHB stars (which they will not all be). We then find the fractional
 absolute residual, $R$, between this distance, which we call $d_*$ and the
 accepted cluster distance $d_0$,
\begin{equation}
R=\frac{|d_*-d_0|}{d_0}.
\end{equation}
We use the absolute residual because the vast majority of sources that have 
distances inconsistent with the cluster distance are placed on the too distant
side (most contaminants will be intrinsically fainter than BHB stars). 
We plot this against the (prior corrected) SVM probability $P(BHB)$. 
Sources with
$P(BHB)>0.5$ are classified as BHB stars for the purposes of this plot. 

In Fig.~\ref{distanceVsProb}, sources which are true BHB stars within the
cluster should appear close to the cluster distance. We would like to see as
many as possible appearing with high probability $P(BHB)$.  Sources which are
in the cluster but are not BHB stars should be assigned distances greater than
the true cluster distance, as they are intrinsically fainter. Ideally, these
sources should of course 
have $P(BHB)<0.5$. BHB stars that are not genuine cluster
members could be in the foreground or the background, so could appear more or
less distant than the true cluster distance. Non-BHB stars in the foreground
or background could appear at greater or lesser distance than the
cluster, but their apparent distance would be greater than their true distance due to their lower luminosity.

 We can see in the figure that there is a clump of apparent BHB stars at the
 cluster distance and with high probability $P(BHB)$. 
The majority of the sources with incompatible
 distances for cluster membership also have $P(BHB<0.5)$. There are a few
 sources with $P(BHB)>0.5$ and incompatible distances. These are probably
 false positive misclassifications, although we cannot rule out the
 possibility that they are simply foreground or background BHB stars.

\section{A catalogue of BHB stars from DR7 photometry}

In Table~\ref{catalogue} we give the basic data for the sample of DR7 sources
classified by us.  The first four columns of this table show various IDs from
SDSS. Column one is the PhotObjId (long) from the SDSS PhotObj table. Columns
two to 4 are the plate ID, MJD, and fiber ID for spectroscopic observations
(where available) from the SDSS SpecObj table. Columns five to eight are the
RA, Dec, l, and b in degrees. Columns nine through thirteen show the SDSS
photometry ($u,g,r,i,z$ psfMags). Columns fourteen through eighteen show the
errors in the photometry in magnitudes. Columns nineteen through twenty-three
show the extinction in each band. Column twenty-four lists the category from
\citeauthor{xue}. This can be either 'BHB', meaning BHB star from the D0.2 fm
method, confirmed by $c_{\gamma}$, $b_{\gamma}$, 'Other', meaning BHB star
from D0.2, fm method, rejected by $c_{\gamma}$, $b_{\gamma}$ method, 'BS',
meaning Blue straggler from D0.2,fm method, 'MS' meaning main sequence star
from D0.2,fm method, or 'None' meaning not present in the \citeauthor{xue}
catalogue.  Column twenty-five shows the raw output probability from the
SVM. Column twenty-six shows the standard deviation of this probability over
ten trials with resampled training set. Column
twenty-seven shows the (2d) prior used. Column twenty-eight shows the
posterior probability calculated from the SVM probability by applying the
prior. Columns twenty-nine and thirty show the assigned distance in kpc and
the fractional error.

In the Appendix, we give all the information needed to directly apply the 
SVM one-class filter and the two-class classifier to new data
for which the SDSS colours, $u-g$, $g-r$, $r-i$ and $i-z$ are available. 

\subsection{A warning on extinction}

The classification is based on dereddened magnitudes, and the dereddening is
performed by the SDSS pipeline based on the map of \cite{schlegel}. This is
expected to work well at high Galactic latitudes, but for sources in the disk
the extinctions may not be reliable.  Furthermore, these maps give the
  line-of-sight extinction to the edge of the Galaxy, so they will
  underestimate extinction in all cases, possibly by a non-negligible amount
  even at high latitudes for nearby BHBs.

The catalogue we present contains
181,022 sources with $|b|<10$ out of the total of 294,652 sources. Out of the
probable BHB stars, 7,231 sources have $P(BHB)>0.5$ and $|b|<10$, and 19,843
sources have $P(BHB)>0.5$ and $|b|>10$.  Users should be aware of this issue
when considering sources at low Galactic latitude.

\begin{figure}[h!]
\begin{center}
 \includegraphics[height=8.cm,angle=270.]{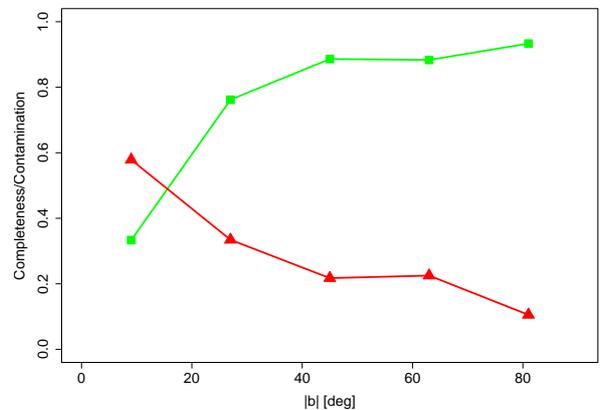}
\end{center}
 \caption{An analysis of classifier performance on the \citeauthor{xue}
testing set as a function of absolute Galactic latitude, $|b|$. 
The completeness (green squares) 
and contamination (red triangles) are plotted. The sample was restricted 
to sources with $g<17.5$. The 2d prior was used, and the contamination
is corrected for the likely class fractions.               
 \label{latitudePerformance}
}
 \end{figure}

To quantify this effect, we calculated the completeness and
contamination in the test sample as a function of absolute Galactic latitude.
The results are shown in Fig.~\ref{latitudePerformance}. 
 From this, we can see that the performance of the classifier holds up
 well for $|b|>30^{\circ}$. Below that, there is some degradation, and the
 performance becomes quite bad in the lowest bin, with $|b|<18^{\circ}$. It is
 difficult to assess the detailed behaviour of the classifier here because of
 the small number of test sources available (there are 63 sources in the first
 bin, of which 18 were BHB stars according to \citeauthor{xue}).

\section{Conclusions}

Starting with a sample of spectroscopically identified BHB stars published by
\cite{xue}, we have trained a number of standard machine learning algorithms
to distinguish BHB stars from other contaminating main sequence stars or other
interlopers, using SDSS colours alone. We have investigated three methods,
with and without the use of probabilistic classification and prior
probabilities, and we find that the support vector machine offers the best
completeness while simultaneously minimizing the contamination in the output
sample. The kernel density estimator was able to provide comparable
contamination, but with a lower completeness. The kNN method was able to match
the completeness of the SVM, but not the contamination.  Adjusting the
  classification thresholds altered this picture in various ways, but the SVM
  generally outperformed the other techniques.

Using the most promising technique (SVM), we have classified a large sample of
DR7 data selected to lie within the colour box of \cite{yanny}.  This sample
comprises 859,341 sources. We used a one-class filter (also based on an SVM), 
to select 294,652 of these as lying in the same colour space
as the available training set.  We have identified 27,074 of these as probable
BHB stars. This includes any already identified by \citeauthor{xue} Because
our classifier relies on a randomly selected subsample of the available training
objects, we ran multiple classifications to quantify the stability of the
output probabilities.  The standard deviations 
of the output probabilities are also provided in the table.

\begin{landscape}
\hspace{-2.cm}
\begin{table}
  \caption{Results of classification of new data. 
} \label{catalogue} 
  \centering
\begin{scriptsize}
\begin{tabular}{@{\extracolsep{-0.10in}}llllllllllllllllllllllllllllll} 
ObjID  & Plate & MJD & Fiber & RA & Dec & l & b & u & g & r & i & z & $\Delta$u & $\Delta$g & $\Delta$ r & $\Delta$ i & $\Delta$ z & u$_{ext}$ & g$_{ext}$ & r$_{ext}$ & i$_{ext}$ & z$_{ext}$ & Type & $P_{SVM}$ & $\Delta P$ & $Q_{g,l}$ & $P'_{SVM}$ & d (kpc) & $\Delta$d/d \\  
587747119446491418 & 0 & 0 & 0 &  0.010639 & -4.960869 & 91.740344 & -64.673367& 23.62 & 22.65 &  22.85 & 22.20 & 22.19 & 0.92 & 0.20 & 0.26 & 0.24 & 0.71 & 0.15 & 0.11 & 0.08 & 0.06 & 0.04 & None & 0.01 & 0.01 & 0.23 & 0.004 & 255.42 & 0.26 \\
587740586797105255 & 0 & 0 & 0 & 0.010989 & 23.874033 & 108.078305 & -37.510853 & 19.70 & 18.43 & 18.39 & 18.40 & 18.39 & 0.03 & 0.01 & 0.01 & 0.02 & 0.03 & 0.57 & 0.42 & 0.30 & 0.23 & 0.16 & None & 0.34 & 0.01 & 0.25 & 0.15 & 31.01 & 0.01 \\
758874300138651868 & 0 & 0 & 0 & 0.031694 & 34.673966 & 111.104952 & -27.016593 & 16.21 & 15.09 & 15.14 & 15.27 & 15.29 & 0.01 & 0.01 & 0.01 & 0.01 & 0.01 & 0.41 & 0.30 & 0.22 & 0.16 & 0.11 & None & 0.19 & 0.02 & 0.43 & 0.15 & 6.99 & 0.01 \\
587740589481525502 & 0 & 0 & 0 & 0.032272 & 26.043768 & 108.765105 & -35.412494 & 21.14 & 19.87 & 20.01 & 20.14 & 20.14 & 0.09 & 0.01 & 0.02 & 0.03 & 0.11 & 0.21 & 0.15 & 0.11 & 0.08 & 0.06 & None & 0.79 & 0.01 & 0.14 & 0.41 & 68.92 & 0.02 \\
758874298530595227 & 0 & 0 & 0 & 0.040556 & 28.786874 & 109.563938 & -32.750327 & 20.98 & 19.99 & 20.03 & 20.07 & 20.14 & 0.06 & 0.02 & 0.02 & 0.02 & 0.08 & 0.30 & 0.22 & 0.16 & 0.12 & 0.08 & None & 0.27 & 0.04 & 0.14 & 0.05 & 69.91 & 0.01 \\
587747072734134392 & 0 & 0 & 0 & 0.044479 & -4.343054 & 92.464250 & -64.139463 & 21.72 & 20.38 & 20.59 & 20.78 & 20.69 & 0.17 & 0.02 & 0.04 & 0.05 & 0.16 & 0.19 & 0.14 & 0.10 & 0.07 & 0.05 & None & 0.99 & 0.38 & 0.23 & 0.96 & 87.27 & 0.04 \\
758874297994576583 & 0 & 0 & 0 & 0.046331 & 26.788068 & 108.999375 & -34.693214 & 23.38 & 22.15 & 22.12 & 21.68 & 21.66 & 0.55 & 0.06 & 0.08 & 0.08 & 0.28 & 0.25 & 0.18 & 0.13 & 0.10 & 0.07 & None & 3.00 & 3.74 & 0.14 & 5.18 & 188.11 & 0.13 \\
587727178449485952 & 0 & 0 & 0 & 0.047126 & -10.668292 & 84.249601 & -69.593383 & 20.00 & 18.82 & 18.82 & 18.92 & 18.88 & 0.05 & 0.02 & 0.02 & 0.01 & 0.05 & 0.17 & 0.12 & 0.09 & 0.07 & 0.05 & None & 0.25 & 0.04 & 0.25 & 0.10 & 42.72 & 0.01 \\
587727225690128568 & 0 & 0 & 0 & 0.049595 & -10.536831 & 84.467652 & -69.486246 & 22.68 & 21.54 & 21.54 & 21.16 & 21.47 & 0.57 & 0.09 & 0.14 & 0.16 & 0.60 & 0.17 & 0.13 & 0.09 & 0.07 & 0.05 & None & 0.29 & 0.30 & 0.25 & 0.12 & 150.65 & 0.14 \\
587747117302546716 & 0 & 0 & 0 & 0.052484 & -1.704633 & 95.003385 & -61.766531 & 21.88 & 20.96 & 20.95 & 20.95 & 20.80 & 0.23 & 0.04 & 0.05 & 0.08 & 0.25 & 0.17 & 0.12 & 0.09 & 0.07 & 0.05 & None & 0.98 & 0.37 & 0.22 & 0.95 & 113.95 & 0.06 \\
587740587334041618 & 0 & 0 & 0 & 0.066353 & 24.353251 & 108.289338 & -37.059177 & 19.45 & 18.07 & 17.97 & 17.99 & 17.99 & 0.03 & 0.01 & 0.01 & 0.02 & 0.02 & 0.49 & 0.36 & 0.26 & 0.20 & 0.14 & None & 0.66 & 0.06 & 0.29 & 0.45 & 27.14 & 0.01 \\
587747119446425658 & 0 & 0 & 0 & 0.067879 & -4.793711 & 92.038001 & -64.551268 & 18.36 & 17.19 & 17.41 & 17.56 & 17.70 & 0.02 & 0.01 & 0.02 & 0.01 & 0.02 & 0.16 & 0.11 & 0.08 & 0.06 & 0.04 & None & 0.74 & 0.01 & 0.50 & 0.74 & 20.34 & 0.01 \\
588015508195639503 & 0 & 0 & 0 & 0.070922 & -0.669070 & 95.923795 & -60.829836 & 21.76 & 20.65 & 20.70 & 20.70 & 20.26 & 0.23 & 0.03 & 0.04 & 0.05 & 0.15 & 0.21 & 0.15 & 0.11 & 0.08 & 0.06 & None & 0.24 & 0.38 & 0.22 & 0.08 & 97.66 & 0.06 \\
758874298530726084 & 0 & 0 & 0 & 0.087467 & 28.430939 & 109.512049 & -33.105980 & 19.83 & 18.70 & 18.68 & 18.56 & 18.42 & 0.03 & 0.01 & 0.02 & 0.02 & 0.03 & 0.31 & 0.23 & 0.16 & 0.12 & 0.08 & None & 0.87 & 0.31 & 0.16 & 0.57 & 38.58 & 0.01 \\
587740525078905365 & 0 & 0 & 0 & 0.088007 & 25.819898 & 108.758217 & -35.641838 & 24.00 & 22.85 & 22.83 & 22.78 & 23.00 & 1.26 & 0.16 & 0.24 & 0.32 & 0.62 & 0.21 & 0.15 & 0.11 & 0.08 & 0.06 & None & 0.46 & 0.11 & 0.14 & 0.13 & 274.36 & 0.34 \\
587731186740822469 & 0 & 0 & 0 & 0.091604 & 0.301938 & 96.745254 & -59.947468 & 23.85 & 22.90 & 23.06 & 23.27 & 23.61 & 0.94 & 0.15 & 0.25 & 0.51 & 0.70 & 0.14 & 0.10 & 0.07 & 0.06 & 0.04 & None & 0.06 & 0.12 & 0.21 & 0.01 & 269.88 & 0.27 \\
587747073270808594 & 0 & 0 & 0 & 0.107236 & -3.878229 & 93.066164 & -63.752046 & 19.59 & 18.48 & 18.60 & 18.77 & 18.88 & 0.04 & 0.01 & 0.01 & 0.02 & 0.03 & 0.19 & 0.14 & 0.10 & 0.08 & 0.05 & None & 0.24 & 0.01 & 0.30 & 0.12 & 36.27 & 0.01 \\
587727179523227719 & 0 & 0 & 0 & 0.110229 & -9.809412 & 85.743476 & -68.913977 & 18.38 & 17.29 & 17.31 & 17.37 & 17.46 & 0.02 & 0.01 & 0.01 & 0.01 & 0.02 & 0.18 & 0.13 & 0.09 & 0.07 & 0.05 & None & 0.13 & 0.00 & 0.51 & 0.14 & 21.04 & 0.01 \\
758874299066679584 & 0 & 0 & 0 & 0.122841 & 30.574360 & 110.133990 & -31.028123 & 19.39 & 18.31 & 18.27 & 18.25 & 18.32 & 0.02 & 0.01 & 0.01 & 0.01 & 0.03 & 0.24 & 0.17 & 0.12 & 0.09 & 0.06 & None & 0.26 & 0.00 & 0.22 & 0.09 & 32.54 & 0.01 \\
587727179523293190 & 0 & 0 & 0 & 0.123358 & -9.954755 & 85.551926 & -69.042703 & 16.21 & 15.12 & 15.27 & 15.45 & 15.58 & 0.01 & 0.01 & 0.01 & 0.01 & 0.01 & 0.19 & 0.14 & 0.10 & 0.07 & 0.05 & None & 0.33 & 0.01 & 0.64 & 0.48 & 7.67 & 0.01 \\
587747119446622618 & 0 & 0 & 0 & 0.220188 & -5.260937 & 91.849680 & -65.035920 & 23.44 & 22.45 & 22.41 & 22.41 & 22.25 & 0.76 & 0.15 & 0.17 & 0.30 & 0.66 & 0.16 & 0.12 & 0.08 & 0.06 & 0.04 & None & 0.99 & 0.33 & 0.23 & 0.99 & 223.05 & 0.20 \\
758874370997223595 & 0 & 0 & 0 & 0.222175 & 27.058341 & 109.262490 & -34.468046 & 16.72 & 15.65 & 15.61 & 15.61 & 15.69 & 0.01 & 0.01 & 0.01 & 0.01 & 0.02 & 0.21 & 0.16 & 0.11 & 0.08 & 0.06 & None & 0.27 & 0.01 & 0.52 & 0.29 & 9.68 & 0.01 \\
588015507658833969 & 685 & 52203 & 188 & 0.226867 & -1.232878 & 95.745605 & -61.408443 & 19.29 & 18.14 & 18.15 & 18.23 & 18.33 & 0.03 & 0.02 & 0.01 & 0.02 & 0.03 & 0.21 & 0.15 & 0.11 & 0.08 & 0.06 & MS & 0.11 & 0.00 & 0.35 & 0.06 & 30.73 & 0.01 \\
587727225690259621 & 0 & 0 & 0 & 0.233561 & -10.446949 & 85.039354 & -69.512947 & 20.47 & 19.35 & 19.40 & 19.49 & 19.39 & 0.08 & 0.02 & 0.01 & 0.02 & 0.08 & 0.21 & 0.15 & 0.11 & 0.08 & 0.06 & None & 0.08 & 0.05 & 0.25 & 0.02 & 53.28 & 0.02 \\
588015508195704946 & 685 & 52203 & 141 & 0.234491 & -0.699505 & 96.206318 & -60.923488 & 20.82 & 19.77 & 19.72 & 19.81 & 19.86 & 0.10 & 0.03 & 0.02 & 0.02 & 0.10 & 0.21 & 0.16 & 0.11 & 0.08 & 0.06 & MS & 0.25 & 0.01 & 0.22 & 0.08 & 65.31 & 0.03 \\
758874297994641467 & 0 & 0 & 0 & 0.240442 & 26.679747 & 109.172876 & -34.839752 & 18.84 & 17.68 & 17.82 & 17.96 & 18.08 & 0.02 & 0.01 & 0.01 & 0.01 & 0.02 & 0.19 & 0.14 & 0.10 & 0.07 & 0.05 & None & 0.39 & 0.01 & 0.30 & 0.22 & 24.98 & 0.01 \\
587747119446753643 & 0 & 0 & 0 & 0.244419 & -5.568110 & 91.561486 & -65.319293 & 22.96 & 21.74 & 21.81 & 21.58 & 21.59 & 0.58 & 0.08 & 0.10 & 0.14 & 0.45 & 0.18 & 0.13 & 0.09 & 0.07 & 0.05 & None & 0.01 & 0.01 & 0.23 & 0.00 & 158.50 & 0.14 \\
758874297994838272 & 0 & 0 & 0 & 0.256721 & 26.225998 & 109.058576 & -35.283989 & 20.71 & 19.61 & 19.61 & 19.65 & 19.67 & 0.06 & 0.01 & 0.01 & 0.02 & 0.06 & 0.21 & 0.15 & 0.11 & 0.08 & 0.06 & None & 0.20 & 0.00 & 0.14 & 0.04 & 60.71 & 0.01 \\
588015509806383417 & 0 & 0 & 0 & 0.287599 & 0.598832 & 97.334501 & -59.750082 & 23.28 & 22.40 & 22.44 & 22.46 & 22.64 & 0.58 & 0.12 & 0.17 & 0.29 & 0.75 & 0.13 & 0.09 & 0.06 & 0.05 & 0.03 & None & 0.02 & 0.04 & 0.21 & 0.00 & 225.32 & 0.15 \\
587747121592730035 & 0 & 0 & 0 & 0.291503 & -1.744949 & 95.428721 & -61.902162 & 22.64 & 21.62 & 21.68 & 21.69 & 21.25 & 0.46 & 0.06 & 0.11 & 0.17 & 0.38 & 0.20 & 0.15 & 0.10 & 0.08 & 0.05 & None & 0.5 & 0.40 & 0.22 & 0.22 & 145.68 & 0.11 \\
587743959426203774 & 0 & 0 & 0 & 0.297441 & 7.314223 & 101.682300 & -53.471631 & 17.74 & 16.54 & 16.51 & 16.55 & 16.59 & 0.01 & 0.01 & 0.01 & 0.02 & 0.02 & 0.32 & 0.23 & 0.17 & 0.12 & 0.09 & None & 0.24 & 0.00 & 0.56 & 0.29 & 14.21 & 0.01 \\
587727225153388676 & 0 & 0 & 0 & 0.307298 & -10.991496 & 84.330154 & -70.004463 & 20.13 & 18.86 & 18.94 & 19.06 & 19.03 & 0.05 & 0.02 & 0.01 & 0.02 & 0.06 & 0.19 & 0.13 & 0.10 & 0.07 & 0.05 & None & 0.71 & 0.03 & 0.25 & 0.45 & 43.32 & 0.01 \\
588015509806383122 & 685 & 52203 & 482 & 0.308759 & 0.537564 & 97.326490 & -59.814775 & 17.38 & 16.13 & 16.20 & 16.33 & 16.39 & 0.01 & 0.02 & 0.01 & 0.01 & 0.02 & 0.13 & 0.09 & 0.07 & 0.05 & 0.03 & BHB & 0.74 & 0.01 & 0.60 & 0.82 & 12.52 & 0.01 \\
\end{tabular}
\end{scriptsize}
\tablefoot{A selection of the first few rows are shown. 
    Some rows were omitted so that some objects with 
    corresponding spectra are shown.    }
\end{table}
\end{landscape}

We used photometric parallaxes derived from colour data presented in 
\cite{sirko} to derive distances
for these objects, using another variant of the support vector machine to
make the fit to the colours. We performed a few simple checks 
on these distances, and on the spatial distribution of the classified 
BHB stars, to demonstrate that our method is reasonable. 

We include along with this work a catalogue of the 294,652 DR7 sources
together with probabilistic identifications as BHB stars, in the hope that
these can be useful for other workers either directly as a ready made BHB
sample, or as prior probabilities for spectroscopic BHB identification
methods.  We also provide, in the appendix, the data and parameters necessary
to apply our classification to new colour data. The accuracy of the
  catalogue, or the classifier, can be estimated by reference to the various
  test results presented in the main body of the paper. In particular,
  Fig.~\ref{magdepcompcontamSvm} gives the estimated performance as a
  function of magnitude, although the reference classifications from the
  \citeauthor{xue} catalogue are unreliable for $g>19$ as seen from
  Figs.~\ref{xueReanalysisFig}
  and~\ref{reanalysisResults}. Figure~\ref{svm_thresholds} gives the expected
  effect of changing the required threshold probability for BHB
  classification, whilst Fig.~\ref{latitudePerformance} can be used to
  estimate the performance as a function of Galactic latitude.

 A general conclusion of this work is that, where reliable training sets
  can be identified, machine learning approaches such as those discussed here
  can probably extract more information than is available with simple colour
  cuts or {\em ad hoc} models. This type of approach is likely to be very
  fruitful in the future for surveys yielding large photometric datasets.

\section*{Acknowledgments}

The authors would like to thank Vivi Tsalmantza, Jelte de Jong, 
Connie Rockosi and Hans-Walter Rix for helpful discussions at various stages 
during the development of this work. We also thank the anonymous referee,
whose comments led to significant improvements in the manuscript. 

Funding for the SDSS and SDSS-II has been provided by the Alfred P. Sloan
Foundation, the Participating Institutions, the National Science Foundation,
the U.S. Department of Energy, the National Aeronautics and Space
Administration, the Japanese Monbukagakusho, the Max Planck Society, and the
Higher Education Funding Council for England. The SDSS Web Site is
http://www.sdss.org/.

The SDSS is managed by the Astrophysical Research Consortium for the
Participating Institutions. The Participating Institutions are the American
Museum of Natural History, Astrophysical Institute Potsdam, University of
Basel, University of Cambridge, Case Western Reserve University, University of
Chicago, Drexel University, Fermilab, the Institute for Advanced Study, the
Japan Participation Group, Johns Hopkins University, the Joint Institute for
Nuclear Astrophysics, the Kavli Institute for Particle Astrophysics and
Cosmology, the Korean Scientist Group, the Chinese Academy of Sciences
(LAMOST), Los Alamos National Laboratory, the Max-Planck-Institute for
Astronomy (MPIA), the Max-Planck-Institute for Astrophysics (MPA), New Mexico
State University, Ohio State University, University of Pittsburgh, University
of Portsmouth, Princeton University, the United States Naval Observatory, and
the University of Washington.

\bibliographystyle{aa}
\bibliography{14381.bib}

\begin{thebibliography}{30}
\expandafter\ifx\csname natexlab\endcsname\relax\def\natexlab#1{#1}\fi
\expandafter\ifx\csname url\endcsname\relax
  \def\url#1{{\tt #1}}\fi
\expandafter\ifx\csname urlprefix\endcsname\relax\def\urlprefix{URL }\fi

\bibitem[{Abazajain et~al.(2009)Abazajain, Adelman-McCarthy, Agueros
  et~al.}]{dr7paper}
Abazajain K., Adelman-McCarthy J., Agueros M., et~al., 2009, ApJS, 182, 543

\bibitem[{Adelman-McCarthy et~al.(2006)Adelman-McCarthy, Agueros, Allam
  et~al.}]{dr4paper}
Adelman-McCarthy J., Agueros M., Allam S., et~al., 2006, ApJ, 162, 38

\bibitem[{Bailer-Jones \& Smith(2010)}]{caljTN53}
Bailer-Jones C., Smith K., 2010, Combining probabilities, Tech. Rep.
  GAIA-C8-TN-MPIA-CBJ-053,
  \urlprefix\url{http://www.mpia-hd.mpg.de/Gaia/publications/probcomb_TN.pdf}

\bibitem[{Bailer-Jones et~al.(2008)Bailer-Jones, Smith, Tiede
  et~al.}]{bailerjones2008}
Bailer-Jones C., Smith K., Tiede C., et~al., 2008, MNRAS, 391, 1838

\bibitem[{Chang \& Lin(2001)}]{changlin}
Chang C.C., Lin C.J., 2001

\bibitem[{Cortes \& Vapnik(1995)}]{cortesvapnik}
Cortes C., Vapnik V., 1995, Machine Learning, 20, 273

\bibitem[{Dorman et~al.(1993)Dorman, Rood, \& O’Connell}]{dorman93}
Dorman B., Rood R.T., O’Connell R.W., 1993, ApJ, 419, 596

\bibitem[{Drucker et~al.(1996)Drucker, Burges, Kaufman, Smola, \&
  Vapnik}]{drucker:support}
Drucker H., Burges C., Kaufman L., Smola A., Vapnik V., 1996, In: NIPS'96,
  155--161

\bibitem[{Gao et~al.(2008)Gao, Zhang, \& Zhao}]{gao}
Gao D., Zhang Y.X., Zhao Y.H., 2008, MNRAS, 386, 1417

\bibitem[{{Harrigan} et~al.(2010){Harrigan}, {Newberg}, {Newberg}
  et~al.}]{harrigan}
{Harrigan} M.J., {Newberg} H.J., {Newberg} L.A., et~al., 2010, \mnras, 621

\bibitem[{Harris(1996)}]{harris}
Harris W., 1996, AJ, 112, 1487

\bibitem[{Hastie et~al.(2001)Hastie, Tibshirani, \& Friedman}]{hastie}
Hastie T., Tibshirani R., Friedman J., 2001, Elements of Statistical Learning,
  Springer-Verlag

\bibitem[{Hayfield \& Racine(2008)}]{np}
Hayfield T., Racine J., 2008, Journal of statistical software, 27, 1

\bibitem[{Huertas-Company et~al.(2009)Huertas-Company, Tasca, Rouan
  et~al.}]{huertas}
Huertas-Company M., Tasca L., Rouan D., et~al., 2009, A\&A, 497, 743

\bibitem[{{Kinman} et~al.(2009){Kinman}, {Morrison}, \& {Brown}}]{kinman}
{Kinman} T.D., {Morrison} H.L., {Brown} W.R., 2009, \aj, 137, 3198

\bibitem[{Li \& Racine(2003)}]{liAndRacine}
Li Q., Racine J., 2003, journal of multivariate analysis, 86, 266

\bibitem[{{Marengo} \& {Sanchez}(2009)}]{marengo}
{Marengo} M., {Sanchez} M.C., 2009, \aj, 138, 63

\bibitem[{Platt(1999)}]{platt}
Platt J., 1999, In: Smola A., Bartlett P., Schoelkopf D. (eds.) Advances in
  large margin classifiers, MIT press

\bibitem[{Richards et~al.(2009{\natexlab{a}})Richards, Deo, Lacy
  et~al.}]{richards2}
Richards G., Deo R., Lacy M., et~al., 2009{\natexlab{a}}, AJ, 137, 3884

\bibitem[{Richards et~al.(2009{\natexlab{b}})Richards, Myers, Gray
  et~al.}]{richards1}
Richards G., Myers A., Gray A., et~al., 2009{\natexlab{b}}, ApJS, 180, 67

\bibitem[{Ruhland et~al.(2010)Ruhland, Bell, Rix, \& Xue}]{ruhland}
Ruhland C., Bell E., Rix H.W., Xue X., 2010, ApJ, Submitted

\bibitem[{Schlegel et~al.(1998)Schlegel, Finkbeiner, \& Davis}]{schlegel}
Schlegel D., Finkbeiner D., Davis M., 1998, ApJ, 500, 525

\bibitem[{S\'ersic(1968)}]{sersic}
S\'ersic J., 1968, Atlas de galaxias australes Cordoba, Argentina: Observatorio
  Astronomio, Tech. rep.

\bibitem[{Sirko et~al.(2004)Sirko, Goodman, Knapp et~al.}]{sirko}
Sirko E., Goodman J., Knapp G., et~al., 2004, ApJ, 127, 899

\bibitem[{Tsalmantza et~al.(2007)Tsalmantza, Kontizas, Bailer-Jones
  et~al.}]{tsalmantza2007}
Tsalmantza P., Kontizas M., Bailer-Jones C., et~al., 2007, A\&A, 470, 761

\bibitem[{Tsalmantza et~al.(2009)Tsalmantza, Kontizas, Rocca-Volmerange
  et~al.}]{tsalmantza2009}
Tsalmantza P., Kontizas M., Rocca-Volmerange B., et~al., 2009, A\&A, 504, 1071

\bibitem[{Vapnik(1995)}]{vapnik}
Vapnik V., 1995, The Nature of statistical learning theory, Springer

\bibitem[{Xue et~al.(2008)Xue, Rix, Zhao et~al.}]{xue}
Xue X., Rix H., Zhao G., et~al., 2008, ApJ, 684, 1143

\bibitem[{{Xue} et~al.(2009){Xue}, {Rix}, \& {Zhao}}]{xue2}
{Xue} X., {Rix} H., {Zhao} G., 2009, Research in Astronomy and Astrophysics, 9,
  1230

\bibitem[{Yanny et~al.(2000)Yanny, Newburg, Kent et~al.}]{yanny}
Yanny B., Newburg H.J., Kent S., et~al., 2000, ApJ, 540, 825

\end{thebibliography}

%\begin{thebibliography}{99}
%\bibitem[\protect\citeauthoryear{Baird}{1981}]{b1} Baird S.R., 1981,
%ApJ, 245, 208
%\end{thebibliography}

\begin{appendix}

\section{Applying the SVM model directly}

The application of the SVM model is mathematically 
straightforward and not excessively laborious. We therefore give here 
the full specification of the SVM classification
so that it can be directly applied to new data. 

The input data are
the four dereddened SDSS colours $u-g$, $g-r$, $r-i$, $i-z$. 
The recipe for applying the model consists of five main steps,
which are listed below. Below, we give detailed
instructions for each step, together with tables containing the 
necessary model data. The steps are:    
\begin{enumerate}
\item Apply the one-class model standardization to the data. 
\item Evaluate the one-class model and reject outliers. 
\item Apply the two-class model standardization to the original data.
\item Apply the two-class model to obtain the decision value, $f$.
\item Apply the probability model to convert $f$ into $P(BHB)$.
\end{enumerate}

This recipe leaves one with an SVM probability implicitly assuming that BHB
stars and non-BHB stars are equal in number in the input sample. 
An appropriate prior should be applied to obtain the posterior probability.

\subsection{Apply the one-class model standardization}

The equation for the standardization is
\begin{equation}
\label{standardizationEqn}
{\mathbf x_s}=\frac{{\mathbf x}-{\bf \mu}}{\bf \sigma},
\end{equation}
where ${\mathbf x}$ are the colours, 
${\mathbf \mu}$ are the means and ${\mathbf \sigma}$ 
the standard deviations of each colour. 
For the one-class model, the standardization is performed using the
  one-class values of ${\mathbf \mu}$ and ${\mathbf \sigma}$ 
given in Table~\ref{standParameters}.

\begin{table}
\caption{Standardization parameters for both one-class and two-class classifiers.  
\label{standParameters}
}
\begin{center}
\begin{tabular}{@{\extracolsep{-0.05in}}ccccc}
          & $u-g$ & $g-r$  & $r-i$  & $i-z$ \\
one-class                                  \\ 
$\mu$     & 1.11589420 & -0.11549278 &  -0.09550841 & -0.09199681 \\
$\sigma$  & 0.09485993 & ~0.08260171 &  ~0.21371742 & ~0.15484431   \\
two-class                                    \\ 
$\mu$     & 1.12810095 & -0.12914688 &  -0.10735233 &  -0.09247437   \\
$\sigma$  & 0.09391075 & ~0.08442376  & ~0.18161768  & ~0.10137230      \\
\end{tabular}
\end{center}
\end{table}

\subsection{Application of one-class SVM model}

The evaluation equation for the SVM model, for either one-class or
  two-class classification, is 
\begin{equation}
f = \sum_{i=1}^{i=N_s} y_i \alpha_i K({\mathbf x},{\mathbf s_i}) - \rho
\label{svmEqn}
\end{equation}
where ${\mathbf x}$ is the colour vector to be classified, ${\mathbf s_i}$ are
the support vectors, $\alpha_i$ their fitted weights, $y_i$ are class labels
for each support vector, and $\rho$ is a constant offset applied to each
result.  The value of $N_s$ is 152 for the one-class model
(Table~\ref{oneClassData}) and $N_s=2645$ for the two class model
(Table~\ref{twoClassData}).  The class labels $y_i$ are set to $+1$ or $-1$
for the two class classifier, and are always set to $+1$ for the one-class
case.

$K$ is a kernel function, in our case an RBF 
kernel, given by
\begin{equation}
K({\mathbf x},{\mathbf s_i})=\exp(-\gamma ||{\mathbf x}-{\mathbf s_i}||^2),
\label{kernEqn}
\end{equation}
where $\gamma$ is a parameter found by tuning (Table~\ref{miscParameters}). 

\begin{table}
\caption{Parameters for one-class, two-class and probability models. 
\label{miscParameters}
}
\begin{tabular}{cc}
\multicolumn{2}{l}{{\bf One-class model}} \\
$\gamma$     &  0.25     \\  
$\rho$       &  2.571136 \\   
\multicolumn{2}{l}{{\bf Two-class model}} \\
$\gamma$     &  0.0625     \\  
$\rho$       &  5.116558   \\
\multicolumn{2}{l}{{\bf Probability model}} \\
$A$       &  -1.162976     \\  
$B$       &   0.006035218  \\
\end{tabular}
\end{table}

The values of the support vectors for the one-class model, corresponding
  to the vectors labeled $s_i$ in Equations~\ref{svmEqn} and~\ref{kernEqn},
  are given in Table~\ref{oneClassData}, which is available in its full form
  as an e-table. The first column in this table gives the product
  $y_i\alpha_i$ for each vector. The value of the parameters $\gamma$ and
  $\rho$ are given in Table~\ref{miscParameters}.

 To apply the one-class model, simply calculate the sum in
  Equation~\ref{svmEqn} over all the support vectors in
  Table~\ref{oneClassData} and subtract the value of $\rho$. Sources with
  $f>0$ are compatible with the training data and are therefore suitable for
  classification with the two-class classifier. Sources with $f<0$ are
  outliers that should be rejected (they cannot be classified with the
  two-class model).

\subsection{Standardization for two-class classification}

 Having rejected sources not compatible with the model, it is now
  necessary to standardize the data for the surviving sources using the
  standardization appropriate for the two-class classifier. The equation for
  this is identical to that used for the one-class standardization,
  Equation~\ref{standardizationEqn} above. The parameters are given in
  Table~\ref{standParameters}. Note that 
this standardization should be performed on
  the {\em original} dereddened colours, 
not on the standardized data used for the 
one-class model.

\subsection{Application of the two-class model}
\label{step4}

 The equations for the two-class model are the same as for the one-class,
  namely~\ref{svmEqn} and~\ref{kernEqn} above. The data for the model should
  be taken from Table~\ref{twoClassData} (support vectors) and from
  Table~\ref{miscParameters} (model parameters).  The decision value $y_i$ in
  Equation~\ref{svmEqn} is now either $-1$ (non-BHB) or $+1$ (BHB), but this
  is of no direct concern to the user since in Table~\ref{twoClassData} the
  value of the product $y_i \alpha_i$ is given.

 Evaluate Equation~\ref{svmEqn} using the two-class data to obtain the
  decision value $f$ for each source. Decision values $f>0$ indicate BHB stars
  (since the class label for BHB stars is $+1$) and decision values $f<0$
  indicate non-BHB stars.

\subsection{Determine the probability of the classification}
\label{step5}

 If only a classification is required, this step is unnecessary. 
If a probability is also required, apply the probability model to 
determine this. 

 Given the value of the decision value $f$ from step~\ref{step4} above,
determine the probability by evaluating
\begin{equation}
P(BHB)=\frac{1}{1+\exp{\left(A f + B\right)}},
\end{equation} 
where A and B are parameters determined by cross validation during training.
The values of these for our model are given in Table~\ref{miscParameters}.

 We note again that this is a 'nominal' probability, assuming equal class
  fractions in reality, no change in class fraction as a function of position,
  magnitude, etc. A prior should be introduced to obtain better posterior
  probabilities, as discussed in Sect.~\ref{sect:priors_general}. The simple
  prior used in this paper of $P(BHB)=0.32$, which roughly accounts for the
  uneven class fractions, is probably the simplest sensible choice for this.

\begin{table}
\caption{Data for the one-class SVM model.
\label{oneClassData}
}
\begin{center}
\begin{tabular}{@{\extracolsep{-0.05in}}ccccc}
$y_{i} \alpha_{i}$ & $u-g$ & $g-r$ & $r-i$ & $i-z$ \\
0.177629  &   -0.125387 &  -3.43222  &  2.042456 &  0.187264 \\
0.853943  &   -2.771393 &  -3.613814 &  0.063207 & -0.607082 \\
0.11659   &    0.665252 &   0.272304 & -1.799065 &  2.331353 \\
0.388386  &    0.412248 &  -0.926218 & -0.278366 & -3.513227 \\
0.255163  &   -0.863317 &  -0.345117 & -1.129957 &  0.309968 \\
1         &    2.562787 &  -3.10535  & -0.250291 & -0.258345 \\
0.382262  &   -0.958194 &   0.70813  & -1.986228 &  2.899666 \\
0.281138  &   -0.515436 &  -2.064209 & -0.535715 & -2.544512 \\
1         &   -1.327159 &  -1.822084 & -2.210824 &  1.937409 \\
\end{tabular}
\tablefoot{Only the
  first ten lines of this table are shown here for illustration, the remainder
  is available in electronic form. There are a total of 152 support vectors in
  the online table. Column one is the product $y_i \alpha_i$, columns two to
  five are the dereddened, standardized colours.}
\end{center}
\end{table}
\begin{table}[!hb]
\caption{Data for the two-class SVM model. 
\label{twoClassData}
}
\begin{center}
\begin{tabular}{@{\extracolsep{-0.05in}}ccccc}
$y_{i} \alpha_{i}$ & $u-g$ & $g-r$ & $r-i$ & $i-z$ \\
1024 &  0.669775 &  0.534765 &  0.034976 &  0.123055 \\ 
1024 & -1.534446 & -1.443351 & -0.4881   & -0.636521 \\
1024 &  0.275784 & -0.507595 & -0.355955 & -0.123561 \\
1024 &  0.584588 &  0.16757  & -0.251339 & -0.172884 \\
1024 & -0.182098 &  0.04912  & -0.256845 &  0.192107 \\
1024 & -0.352472 & -0.957706 & -0.543161 & -0.212342 \\
1024 & -0.586737 &  0.85458  &  0.657162 & -0.291259 \\
1024 & -0.97008  &  1.340226 &  0.18364  &  0.724797 \\
1024 & -0.064965 & -0.720806 & -0.229315 & -0.498417 \\
\end{tabular}
\tablefoot{Only the first ten lines are shown here for illustration. 
The remainder is available in electronic form. 
There are a total of 2,645 support vectors 
in the full online table. The columns have the same 
meaning as for Table~\ref{oneClassData} above. 
Note that column 1, which gives the product $y_i \alpha_i$, 
always has the same value for the first ten instances. 
This is not the case for all  the support vectors in the full version.}
\end{center}
\end{table}

\end{appendix}

\end{document}